\documentclass[aps,twocolumn,amsmath,amssymb,showpacs,eqsecnum,floatfix,rmp]{revtex4}
\usepackage{epsf}
\usepackage{dcolumn}
\usepackage{bm}
\newcounter{saveeqn}%

\begin{document}
\title
{\bf Weakly Interacting, Dilute Bose Gases in 2D}
\author{Anna Posazhennikova}
\email[Email:]{anna@tfp.physik.uni-karlsruhe.de} 
\affiliation{Institut f\"ur Theoretische Festk\"orperphysik and Institut f\"ur Theorie der Kondensierten Materie, Universit\"at Karlsruhe, 76128 Karlsruhe, Germany}
%\date {June 14,2006}
\begin{abstract}
This article surveys a number of theoretical problems and open questions in the field of two-dimensional dilute Bose gases with weak repulsive interactions. 
In contrast to three dimensions, in two dimensions 
the formation of long-range order is prohibited by the Bogoliubov-Hohenberg theorem, and Bose-Einstein condensation is not expected to be realized.
Nevertheless, the first experimental indications supporting the formation of a condensate in low dimensional systems have been recently obtained. This unexpected behaviour appears to be due to the non-uniformity introduced into a system by the external trapping potential. 
Theoretical predictions, made for homogeneous systems, require therefore careful reexamination. 
  We survey a number of popular theoretical treatments of the {\it dilute} weakly interacting Bose gas and discuss their regions of applicability. The possibility of  Bose-Einstein condensation in a two-dimensional gas, the validity of the perturbative $t$-matrix approximation and the diluteness condition are issues that we discuss in detail.
\end{abstract}


\maketitle

\tableofcontents

\section{Introduction}

\subsection{Revival of interest in low-dimensional systems}

Low-dimensional systems are interesting in general, as their low-temperature physics is governed by strong long-range  fluctuations. 
These fluctuations inhibit the formation of the true long-range order (LRO), which is a key concept of phase transition theory in 3D. Thus, a 2D uniform system of interacting bosons does not undergo Bose-Einstein condensation at finite temperatures. However, this system turns superfluid below a certain temperature $T_{KT}$, identified by Berezinskii, and Kosterlitz and Thouless (BKT) in 1971-73, signalling the presence of a so-called topological order. The elementary excitations of the superfluid phase are pairs of vortices with opposite winding numbers. 
  
The experimental realization of such a system was for many years restricted to films of superfluid $^4$He on surfaces, which is also an example of a  strongly-interacting system. The breakthroughs in experimental physics at the end of the last century have changed the situation drastically. The combination of laser cooling (S. Chu, C. Cohen-Tannoudji, W. D. Phillips, Nobel Prize for Physics, 1997) with evaporative cooling and magneto-optical traps provided experimental systems of cold atoms, which were primarily used to observe the long-awaited phenomenon of Bose-Einstein condensation (E. A. Cornell, W. Ketterle, E. Wieman, Nobel Prize for Physics 2001). 
The full {\it tunability} of magnetic and optical traps opens an extraordinary opportunity to study in practice not only 1D and 2D Bose systems, but also dimensional crossovers under the influence of the number of particles, size and shape of the system, interaction strength and temperature. These new developments have triggered a revival of theoretical interest in low-dimensional systems, when the old theoretical predictions are to be tested or carefully revised in order to address {\it finite-size} experimental systems, and a large field of new phenomena are to be explained.  

While the first experimental indications of the BKT transition in weakly-interacting Bose system have been recently obtained \cite{Stock:2005}, many questions remain unanswered. One of the most interesting is, whether topological order survives under some conditions in the inhomogeneous trapped system, or is it dominated by the true LRO and Bose-Einstein condensation prevails? Can we control and directly observe the formation of vortex pairs in 2D quantum gases? These and other problems serve as the main motivation for this Colloquium.

In the next section we present a succinct overview of the history of work with dilute Bose systems, outlining some of the important theoretical problems  relevant to weakly-interacting Bose gases.  

\subsection{Historical overview}

The condensation of conserved particles that obey the same statistics as photons was predicted by Einstein in 1924 even before the concept of Fermi statistics was introduced (1926). Einstein's prediction was preceded by an ingenious conjecture of  Bose, who realized that black body radiation can be treated as a gas of {\it indistinguishable} photons. Einstein generalized ideas of Bose to material particles and published two famous papers, in which he developed what we now call Bose-Einstein statistics \cite{Einstein:1924,Einstein:1925}.

The ideal gas of Bose particles is remarkably the only example of a {\it non-interacting} system in condensed  matter physics that undergoes a phase transition upon decreasing the temperature. 
However, experimental realization of ideal Bose-Einstein condensates is extraordinarily difficult, since realistic systems always involve interactions. Largely for this reason Einstein's ideas did not receive a wide recognition in the scientific community for many years as being devoid of any practical significance. The condensation phenomenon did not even appear in the textbooks, until in 1938 F. London recognized the analogy between superfluidity of liquid $^4$He, discovered by \textcite{Kapitza:1938}, and \textcite{Allen:1938}  and an ideal Bose gas and emphasized that Einstein's statement was ``erroneously discredited'' \cite{London:1938}. 

In support of London's phenomenological ideas, the first {\it microscopic} theory of superfluidity in a system of weakly-interacting Bose particles was introduced in a brilliant paper by \textcite{Bogoliubov:1947}. 
Subsequent discussion about the connection between superfluidity and Bose Einstein condensation led \textcite{Penrose:1956} to formulate the  generalized criterion for BE condensation. This line of research culminated in a paper of C. N. Yang, who in 1962 extended this criterion to superfluidity and superconductivity and proposed  the concept of off-diagonal long-range order (ODLRO) \cite{Yang:1962}. The condensed phase is characterized then by a non-vanishing asymptotic of a one-body density matrix at large distances. 

During the decades which followed the work of Bogoliubov, successful field-theoretical approaches were developed and many important predictions about the thermodynamics of the interacting Bose system were made. However, apart from the successful observation of superfluidity in liquid Helium systems, the quest to create  Bose-Einstein condensates (BEC) proved unrewarding for several decades. Finally, in 1995
 Bose-Einstein condensates were
realized in a fascinating series of experiments on rubidium and sodium vapours \cite{Ketterle:1999,Ketterle:2001,Cornell:2002}). The importance of this experimental achievement was recognized in the 2001 Nobel Prize for Physics, shared by E. A. Cornell, W. Ketterle, and E. Wieman.

The experimental realization of BEC has offered a unique opportunity to probe and control many interesting phenomena, not accessible or unstudied in the field of superfluidity, such as dimensional transitions, the crossover from Bose Einstein condensation to BCS pair condensation, interference effects, and disorder effects. Exotic links to cosmology \cite{Fedichev:2003}, quantum optics \cite{Recati:2005} (two-state atomic quantum dots within a condensate), and even wetting phenomena \cite{Indekeu:2004} have been recently proposed. The growing interest in Bose systems has resulted in more than 600 studies per year during last decade and the list of references related to BEC now exceeds 200 pages!

The actual observation of condensation was hindered by enormous technical difficulties, so that even 15 years ago researchers dared not to believe that nature would ever provide them with the ``right'' system. The main problem to overcome is the condensation of most systems into a solid or liquid upon cooling to low temperatures, which by-passes the BEC transition. In particular, the formation of clusters or molecules is driven by three-body collisions. The hard task for an experimentalist was therefore the creation of a gaseous system, in which three-body collisions occur much less frequently than  two-body interactions.

The gas in which the two-body interactions prevail is called dilute.
 Diluteness implies a very low density of the gas, so that the characteristic range $a_s$ of the potential between the Bose particles is small compared to the mean particle distance, proportional to  $n^{-1/3}$ in three dimensions ($n=N/V$ being the density of the gas). The diluteness condition  is therefore equivalent to the requirement that the {\it gas parameter} $n^{1/3}a_s$  be small
\begin{equation}
n^{1/3}a_s\ll 1.
\label{dilute3D}
\end{equation}

Ultra low density of the system leads to extremely low condensation temperatures (in the nanokelvin range), realization of which was another technical obstacle for the experimentalists. At low temperature the thermal velocity of the particles $v_T$, which is proportional to the inverse De Broglie wave length
\begin{equation}
\lambda_T=\sqrt \frac{2\pi \hbar^2}{m k_B T},
\end{equation}
becomes very small ($v_T=\hbar/m\lambda_T \sim$ 1 mm/sec) and at temperatures of the order of a few nK all the particles ``jump'' into a coherent ground state. Sufficient {\it  diluteness} of the gas is therefore one of the crucial conditions for BEC to be observed in the experiment. 

In order to reach the required regime of temperature and density, various cooling and trapping techniques have been developed \cite{Ketterle:1999}. Before being cooled atoms are confined in an external potential created by an applied magnetic field.
The finite extent of the condensate cloud and its inherent inhomogeneity introduce a number of important differences between BEC in a trap and uniform gas. 
For example, a trapped gas of Bose atoms exhibits a BEC transition not only 
 in momentum space, but in coordinate space as well \cite{Dalfovo:1999}. In practice however, condensates are so small that the literal observation of their size and shape is limited by the resolution of existing experimental equipment. Nevertheless real space Bose condensates provide a novel resource for exploring many interesting phenomena, such as quantum interference effects and frequency dependent collective excitations. 

The effect of a magnetic trap becomes more dramatic for lower dimensionality of the system. 
For example, in 2D a noninteracting trapped gas undergoes a BEC phase transition at finite temperature \cite{Widom:1968,Bagnato:1991,Li:1999} in contrast to the 2D uniform case, where condensation is possible only at zero temperature. This difference arises due to modification of the density of states of the gas in the presence of a trap.

The description of an {\it interacting} system in a 2D harmonic potential is not trivial. In the case of a uniform gas, long range order does not develop because of the preponderance of long wavelength phase fluctuations, inherent to low-dimensional systems. This can be also seen as an infrared divergence of the integral $\int N({\bf p})\frac{d^2p}{(2\pi\hbar)^2}$, where $N({\bf p})$ is the number of particles out of the condensate with momentum $\bf p$. This divergence, on the other hand, is a consequence of the fact, that the energy of the system depends only on the phase gradient, and not on the phase itself, because the latter is not a well-defined quantity \cite{Lifshitz:2004}. The absence of long-range order in 2D systems with a continuous symmetry is often referred to as the Bogoliubov $k^{-2}$, or Hohenberg-Mermin-Wagner (BHMW)  theorem (see works by \textcite{Bogoliubov:1961,Bogoliubov:1991}, \textcite{Wagner:1966}, \textcite{Mermin:1966}, and \textcite{Hohenberg:1967}), and we discuss this issue in more detail in Chapter \ref{LRO2D}. \textcite{Fisher:1988} pointed out that a consequence of the long-wavelength phase fluctuations is a drastic modification of  the diluteness condition, so that the conventional low-density requirement  for weakly-interacting  2D Bose  gas, $n^{1/2}a_s\ll 1$ is replaced by an inequality
\begin{equation}
\ln \ln\frac{1}{na_s^2}\gg 1.
\label{ineqFH}
\end{equation}
Taken literally, condition (\ref{ineqFH}) rules out the possibility of experimental realization of a 2D {\it dilute} Bose system. However, this condition does not work {\it away} from the transition. One can show from the analysis of quantum fluctuations  (see \textcite{Petrov:2004} for review) that in this case
 the diluteness criterion amounts to $1/\ln(1/na^2)\ll 1$, previously derived by \textcite{Schick:1971}.

It is also intuitively clear that the trapping potential introduces a lower bound for the momentum of excitations and thus prevents the establishment of the long-range thermal fluctuations which destroy the condensate. Based on these arguments, \textcite{Petrov:2000} showed the existence of a true condensate in a quasi-2D system in a wide parameter range. 

More generally, the BHMW approach is not really suitable for a proper analysis of an inhomogeneous system, such as trapped atomic vapour, as pointed out by \textcite{Fischer:2002,Fischer:2005}. 
In his work \textcite{Fischer:2002,Fischer:2005} obtained a geometrical equivalent of the BHMW theorem, independently of the Hamiltonian of the system  and showed that in the marginal $d=2$ case true condensation  is still possible in an appropriately defined thermodynamic limit.

In support of theoretical estimations, the first experimental confirmations of macroscopic occupation of the harmonic oscillator ground state \cite{Goerlitz:2001,Rychtarik:2004} became known in sodium atom vapours, confined to optical and magnetic traps. A rapid progress in experimental techniques made it possible to increase the aspect ratio (anisotropy) of the trap from 79 \cite{Goerlitz:2001} to 700 \cite{Smith:2005}. This large anisotropy of the new traps is sufficient to confine condensates with $\sim 10^5$ atoms in a quasi-2D regime \cite{Smith:2005}. 
 Signs of local coherence were also observed in a two-dimensional gas of hydrogen atoms, absorbed on liquid $^4$He surface \cite{Safonov:1998}. Quasi-2D condensate have been also recently created by \textcite{Stock:2005} and interesting phase defects have been measured. The crossover from 3D condensates to two- and ultimately 1D can be observed by changing the aspect ratio of the trap.

As indicated in previous section, the recent progress in laser-based trapping techniques 
and creation of optical lattices has led to a new generation of remarkable experiments. With controllable interparticle interaction it is now possible to observe 
 the transition from the superfluid state to a Mott insulator \cite{Bloch:2004}. 
Optical lattices provide a way to investigate various intriguing aspects of low-dimensional systems as well. Interest in 2D configurations of Bose particles has arisen in the context of high-temperature superconductivity and the fractional quantum Hall effect. All in all, ultracold atomic gases have the potential to impact a very broad range of physics. 
 
In this Colloquium we discuss a number of selected issues related to two-dimensional weakly-interacting neutral Bose gases. If necessary, 3D problems are mentioned. We attempt to cover many references and otherwise refer the reader to numerous resources, such as several excellent 
 theoretical reviews  \cite{Dalfovo:1999,Castin:2001,Leggett:2001,Fetter:2002,Petrov:2004,Yukalov:2004} and books \cite{Pines:1962,Griffin:1993,Pethick:2002,Pitaevskii:2003}, a Resource Letter for BEC \cite{Hall:2003} and BEC web-sites. 
Though a certain level of subjectivity is unavoidable, we aim to provide the necessary information about the field to those 
 who feel lost after a preliminary contact with current literature but want to learn more about the main problems in the fascinating area of Bose-Einstein condensates in 2D.

\section{Ideal Bose gas} \label{Ideal}

Consider a macroscopic system of noninteracting Bose particles at finite temperature in the grand-canonical ensemble. The total number of particles in such a system is defined by the equation
\begin{equation}
N=\sum_k n_B(\epsilon_k)=\int \rho(\epsilon)n_B(\epsilon)d\epsilon,
\label{numberofparticles}
\end{equation}
where $n_B(\epsilon_k)=1/(\exp \beta (\epsilon_k-\mu)-1)$ is the Bose-Einstein distribution function, $\beta=1/k_BT$ and $\rho(\epsilon)$ is the density of states.

The chemical potential $\mu$ of the Bose gas, being negative, increases as the temperature drops and vanishes at the critical temperature $T_c$, indicating the phase transition to a condensed state. The transition temperature is therefore defined by Eq.(\ref{numberofparticles}) with $\mu=0$.
In $d$ dimensions the density of states $\rho(\epsilon)=dN_{\epsilon}/d\epsilon\sim \epsilon^{(d-2)/2} $, and the particle density is proportional to the integral
\begin{equation}
n\equiv \frac{N}{V} \sim \int\frac{\epsilon^{(d-2)/2} d\epsilon}{exp(\epsilon/T_c)-1}. 
\label{tceq3d}
\end{equation}
In 3D this integral converges and the Bose-Einstein condensation temperature has a finite value, $T_c^{3D}\sim n^{2/3}$. This result can be also understood as a temperature scale at which the thermal wave length becomes comparable with the average interparticle spacing $\lambda_T\sim l\sim n^{-1/3}$. As $\lambda$ is proportional to $T^{-1/2}$, $T_c$ is proportional to $n^{2/3}$.

One can also calculate the 
and the number of particles occupying the ground state
\begin{equation}
N_0=N\left(1-\frac{T}{T_0}\right)^{3/2}.
\end{equation}
It is readily seen that $N_0$ increases as the temperature decreases. This phenomenon of macroscopic occupation by  particles of the state with minimal energy at low temperatures is referred to as Bose-Einstein condensation. Note that the actual condensation occurs in momentum space. 

In 2D, a constant density of states leads to an infrared divergent integral in expression (\ref{tceq3d}) and condensation is not possible at any finite temperature. 

Let us discuss  how this picture  changes in the presence of a trap. 
The general treatment of this problem was considered by \textcite{Bagnato:1991}. They studied the possibility of the Bose-Einstein condensation of an ideal gas, confined by one- or two-dimensional power-law trap: $V_{ext}\sim x^{\eta}$. \textcite{Bagnato:1991} showed that a two-dimensional system undergoes BEC for any finite value of $\eta$, moreover, $T_c^{2D}$ has a broad maximum in the vicinity of $\eta=2$, i.e. for a trapping potential close to parabolic. (A one-dimensional system displays BEC only for $\eta <2$.)

Practically the confining trap is well approximated by a harmonic potential
\begin{equation}
V_{ext}({\bf r})=\frac{m}{2}(\omega_x^2x^2+\omega_y^2y^2+\omega_z^2z^2).
\end{equation}

For non-interacting particles we can write the many-body
Hamiltonian as  a sum of one-particle Hamiltonians $H_{MB}=\sum_{i=1}^N H_{SP}(i)$,
whose eigenvalues are
\begin{equation}
\epsilon_{n_xn_yn_z}=(n_x+\frac{1}{2})\hbar\omega_x+
(n_y+\frac{1}{2})\hbar\omega_y+
(n_z+\frac{1}{2})\hbar\omega_z.
\label{spectrum}
\end{equation}
 The lowest energy of the system in the trap is $\epsilon_{000}=\frac{3}{2}\hbar \overline \omega$, where for the sake of simplicity we introduced the average frequency $\overline \omega=(\omega_x+\omega_y+\omega_z)/3$.

Note, that in the ground state all $N$ particles occupy the level $\epsilon_{000}$ and the wave function of the ``cloud'' of these particles is easy to find 
\begin{eqnarray} 
&&\phi({\bf r_1..r_N})=\prod_i \varphi_0({\bf r_i}) \nonumber \\
&&\varphi_0({\bf r_i})=\left(\frac{m\omega_{ho}}{\pi\hbar}\right)^{3/4}
\exp\left(-\frac{m}{\hbar}(\omega_xx^2+\omega_yy^2+\omega_zz^2)\right)
\label{spwf}
\nonumber
\\
\end{eqnarray}
where
\begin{equation}
\omega_{h0}=(\omega_x\omega_y\omega_z)^{1/3}.
\end{equation}
In this case the density distribution of the particles is position dependent
\begin{equation}
n({\bf r})=N|\varphi_0({\bf r})|^2
\end{equation}
and the first important length scale appearing in the problem is the size of the cloud
\begin{equation}
a_{h0}=\sqrt{\frac{\hbar}{m\omega_{h0}}}
\label{trapsize}
\end{equation}
which is just the  average width of the Gaussian distribution (\ref{spwf}) (Fig.1). Experimentally $a_{h0}$ is usually of order of 1$\mu m$.

\begin{figure}[!ht]
\begin{center}
\epsfxsize=0.45\textwidth{\epsfbox{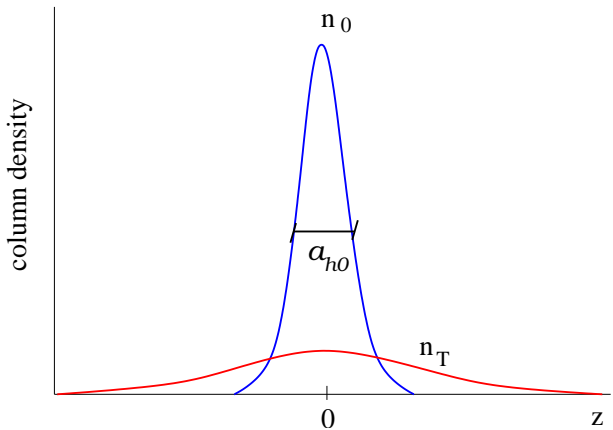}}
\end{center}
\caption{Column density of a cloud of trapped noninteracting bosons along the $z$-direction. The total density is a superposition of a condensate density $n_0$ and a thermal distribution of noncondensed particles $n_T$.}
\label{tmatrix}
\end{figure}

Since we are mostly interested in low-dimensional effects, it is instructive to mention the experimental realization of a two-dimensional atomic trap. An axially symmetric harmonic potential can be written in the form, $V_{ext}(r)=\frac{1}{2}m\omega_{\perp}^2r_{\perp}^2+\frac{1}{2}m\omega_z^2z^2=\frac{1}{2}m\omega_{\perp}^2(r_{\perp}^2+\lambda^2z^2)$, where $\lambda=\omega_z/\omega_{\perp}$ characterizes the degree of anisotropy. 
For $k_BT\ll \hbar\omega_z$ and $k_BT>\hbar\omega_{\perp}$ the motion of atoms along the $z$ direction is frozen  (particles only  undergo zero point oscillations), and kinematically the gas can be considered as two dimensional. 
Thus by making one dimension of the trap very narrow,  oscillator states become widely separated, 
and an effective 2D system is realized.

At finite temperature only a fraction of the particles $N_0$ occupies the lowest energy level and the others are thermally distributed over higher energy levels. However, we still can treat $N_0$ as a macroscopic number. Thermal excitations will cause the size of the atomic cloud to grow with temperature. In the {\it semiclassical} approximation $k_BT>>\hbar \omega_{h0}$, where 
 the relevant excitation energies are much larger than the interlevel spacing, it can be shown that the size of the cloud increases as a square root of temperature $R_T=a_{h0}\sqrt \frac{k_BT}{\hbar\omega_{h0}}$. The important conclusion of this short discussion is that in harmonic traps, Bose condensation manifests itself as sharp peak in the central region of the density distribution in real space. The appearance of such a peak in both coordinate and momentum space is a peculiar feature of the trapped condensates, with significant impact on both theory and experiment. This is very different from the uniform gas discussed above, where the condensation cannot be revealed in  real space, for the condensate and uncondensed particles occupy the same volume.

The total number of particles in the trap is defined  by 
\begin{equation}
N=\sum_{n_xn_yn_z}\frac{1}{\exp(\frac{\epsilon_{n_xn_yn_z}-\mu}{T})-1}
\label{nrofparticlestrap}
\end{equation}
which is derived from Eq.(\ref{numberofparticles}) with a  discrete energy spectrum (\ref{spectrum}).
Note, that in this case the chemical potential at the transition point acquires a non-zero value of the lowest energy level: $\mu(T\rightarrow T_c)\rightarrow \mu_c=(3/2)\hbar \overline \omega$.

In the semiclassical approximation we can simplify (\ref{nrofparticlestrap}) by replacing the summation with integration and a straightforward  solution for $\mu=\mu_c$ gives the Bose-condensation temperatures for the trapped gas in three and two dimensions
\begin{equation}
T_c^{3D}=\frac{\hbar}{(\zeta(3))^{1/3}}\omega_{ho}N^{1/3}
\label{tc3d}
\end{equation}
\begin{equation}
T_c^{2D}=\frac{\hbar\sqrt{6}}{\pi}\omega_{ho}N^{1/2}
\label{tc2d}
\end{equation}

The two-dimensional condensation temperature is now finite (nonzero). This is related to the density-of-states effect of the gas in the trap. Indeed, in the semiclassical approximation we can introduce a coordinate system defined by the three variables $\epsilon_{x,y,z}=\hbar n_{x,y,z}\omega_{x,y,z}$, in terms of which the surface of constant energy (\ref{spectrum}) is the plane $\epsilon=\epsilon_x+\epsilon_y+\epsilon_z$. Then the number of states $N(\epsilon)$ is proportional to the volume in the first octant bounded by this plane
\begin{equation}
N_{\epsilon}=\frac{1}{\hbar^3 \omega_{h0}^3}\int_0^{\epsilon}d\epsilon_x\int_0^{\epsilon-\epsilon_x}d\epsilon_y\int_0^{\epsilon-\epsilon_x-\epsilon_y}d\epsilon_z=\frac{\epsilon^3}{6\hbar^3\omega_{h0}^3}
\end{equation}
 The density of states $\rho=dN_{\epsilon}/d\epsilon$ is then quadratic in energy  $\rho^{3D}\sim\epsilon^2$ in three dimensions and linear in  energy in two dimensions $\rho^{2D}\sim \epsilon$, in contrast to the constant density of states of a uniform 2D gas, and the integral in the Eq.(\ref{nrofparticlestrap}) for $\mu=\mu_c$ is not infra-red divergent until $d=1$.

It is now straightforward to calculate the condensate fraction (e.g. 3D)
\begin{equation}
\frac{N_0}{N}=1-\left(\frac{T}{T_{c}^{3D}}\right)^3 
\end{equation}
and total energy of the system and correspondingly all the interesting  thermodynamic quantities. In 2D the condensate fraction is \cite{Bagnato:1991,Petrov:2004}
\begin{equation}
\frac{N_0}{N}\approx 1-\left(\frac{T}{T_{c}^{2D}}\right)^2. 
\label{condfrac_2D}
\end{equation}
Sign "$\approx $" in the expression \eqref{condfrac_2D} is related to the fact, that at $T=T_c^{2D}$ the condensate fraction is not exactly zero, because there is a small correction to the result due to the finite number of particles in the system \cite{Petrov:2004}. One should be therefore careful with the word ``phase transition" in the context of trapped gases, because they are finite size systems and the phase transition notion is strictly defined only in thermodynamic limit. It is better to say that at $T_c$ there is a sharp crossover to the BEC state in the system. Note also, that at $T_c^{2D}$ the de-Broglie wavelength $\lambda_T$ becomes comparable with the mean interparticle separation $\sim \sqrt{T_c/(Nm\omega_{ho}^2})$.

We end the section by remarking on the proper definition of the thermodynamic limit in the trapped case.
It is well known that the transition temperature should be well-defined in the thermodynamic limit. The usual definition  when the ratio $N/V$ is kept constant while the number of particles $N$ and the volume $V$ tend to infinity is apparently not suitable for the inhomogeneous situation. The appropriately defined limit is then obtained by letting $N\rightarrow \infty$ and $\omega_{h0}\rightarrow 0$, while keeping the product $N\omega_{h0}^3$ (or $N\omega_{h0}^2$ in 2D) constant. In this case the temperatures (\ref{tc3d}) and (\ref{tc2d}) are  well-defined.

A comprehensive survey of various issues related to the behaviour of the ideal Bose gas in a harmonic potential can be found in the paper by \textcite{Mullin:1997}. 

The ideal-gas results are summarized in the Table \ref{tab_ideal}.

\section{Ground state of a weakly interacting Bose gas} \label{Ground}

\subsection{Bogoliubov approximation} \label{RPA_Bog}

In his seminal paper "On the theory of superfluidity'' \cite{Bogoliubov:1947}, published in 1947, Bogoliubov introduced the microscopic description of the ground state of a uniform, weakly interacting Bose gas. 
The assumption about the uniformity of the unperturbed ground state is crucial to his results. To assure a uniform Bose gas, Bogoliubov considered the case of repulsive interactions and made use of periodic boundary conditions. The gas is also assumed to be dilute ($na^3\ll 1$), which permits to simplify the many-body problem and account for interactions in a rather fundamental way. In contrast to the uniform case, the nonuniform ground state 
 is very ``sensitive'' to the introduction of any interactions and makes the solution of the many body problem highly nontrivial.

The standard Hamiltonian of an interacting Bose gas is
\begin{eqnarray}
H&=&\frac{1}{2}\int \bigtriangledown \Psi^{+}(r)\bigtriangledown\Psi(r)dr+
\nonumber \\
&+&\frac{1}{2}\int \Psi^+(r)\Psi^+(r')U(r-r')\Psi(r')\Psi(r)dr dr' \qquad
\end{eqnarray}
where $U(r-r')$ is the interaction between particles. In momentum space this Hamiltonian reads
\begin{equation}
H=\sum_p\epsilon_p a_p^+ a_p+\frac{1}{2 V}\sum_{pp'q} U_q a^+_{p}a^+_{p'}a_{p'-q} a_{p+q},
\label{Ham}
\end{equation}
 $U_q=\int e^{-iqr}U(r)dr$ is a Fourier component of the interaction, the bosonic field operator $\Psi(x)=1/\sqrt{V}\sum_p e^{ipx}a_p$ (here $x$ is a four vector), and  the boson creation and annihilation operators satisfy the usual commutation relations $[a_p,a_{p'}^+]=\delta_{pp'}$.

 Without interactions all $N$ particles of the system occupy the state with zero energy and
zero momentum. 
 The number of condensed particles  $N_0$ in this case is equal to the total number of particles 
 $N$. 
When we switch on the interaction, two particles can scatter out of the condensate and occupy one of the the many zero-total-momentum states
with separate momenta ${\bf k}$ and ${\bf -k}$ (in the lowest order
perturbation theory) and $N_0$ naturally decreases.

For a dilute weakly interacting Bose gas one can assume that the total depletion of the condensate is small ($\delta N/N_0\ll 1$) and most of the particles remain in the condensate $N_0\gg 1$. 
The key observation of Bogoliubov is that in this case the second-quantized
condensate operators can be simply replaced by the ``c''- number $\sqrt N_0$
\begin{equation}
\hat a_0, \hat a_0^+ \sim \sqrt{N_0}.
\label{prescription}
\end{equation} 
The drawback of this prescription is that it leads to a Hamiltonian which no longer conserves
 the number of particles. This problem can be partly resolved by working in the grand-canonical ensemble, in which additional terms  $-\mu N_p$ ($N_p=\sum_{p\neq 0}a_p^+a_p$) are introduced into the Hamiltonian (\ref{Ham}). This secures  the conservation of particles on the average.
It is also worth mentioning that the Bogoliubov approximation is equivalent to the neglect of any dynamics in the condensed state. 

In the weak coupling limit the Hamiltonian (\ref{Ham}) can be diagonalized by applying the Bogoliubov canonical transformation
\begin{eqnarray}
a_k=u_k \alpha_k-v_k\alpha^+_{-k} \nonumber \\
a_k^+=u_k\alpha_k^+-v_k \alpha_{-k}
\end{eqnarray}
and the resultant Hamiltonian describes the system of {\it non-interacting} quasiparticles with the spectrum
\begin{equation}
\xi_k=\sqrt{n_0U_0\frac{k^2}{m}+\frac{k^4}{4m^2}},
\label{quasi}
\end{equation}
where $n_0=N_0/V$ is the density of condensed particles.

From this dispersion relation (\ref{quasi}) it follows that in the long wavelength limit the Bogoliubov quasiparticles behave as ``phonons'' with a sound velocity $s=\sqrt{\frac{n_0U_0}{m}}$, and all of the low temperature thermodynamics of a Bose-condensed system is governed by this phonon spectrum. In the opposite limit of short wavelength the quasiparticles behave as {\it free} particles with an energy $\frac{k^2}{2m}$. By equating the kinetic energy and the ``Hartree'' interaction energy $n_0U_0$ one can straightforwardly find the ``transition'' wave vector $k_c=\sqrt{2mn_0U_0}\sim \sqrt{2}ms$, which separates the phonon-like behavior of elementary excitations from the free particle one. $k_c$ introduces an important length scale into the system (Fig.2)
\begin{equation}
\lambda_c=\frac{\hbar}{k_c}=\hbar/\sqrt{2mn_0U_0},
\label{healing}
\end{equation} 
over which the coherence effects are important in the interaction between particles.
It is usually called the {\it healing length} (as in the context of trapped condensates), or sometimes the correlation or coherence length, and it refers to correlations between excitations in the system. These correlations are distinct from the long-range correlations, which lead to condensation in the $k=0$ mode.

\begin{figure}[!ht]
\begin{center}
\epsfxsize=0.45\textwidth{\epsfbox{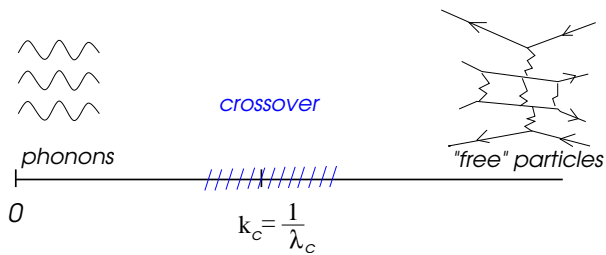}}
\end{center}
\caption{Length scale in the Bogoliubov problem: correlation length $\lambda_c$. $k_c\sim 1/\lambda_c$ separates the ``free'' particle behaviour from the linear dispersion region. Zigzag lines denote the residual interaction between the particles.}
\label{tmatrix}
\end{figure}

One should also mention  that the Bogoliubov canonical transformation is equivalent 
to a
summation over the most divergent terms in the perturbation-series
expansion for the ground state energy. The summation of such series is also
equivalent to making the random-phase approximation (RPA).

It was important in the theory of superfluidity, that the low-lying Bogoliubov quasiparticles follow a linear dispersion. This kind of behavior is fully consistent with the Landau criterion for superfluidity, i.e.,  that no excitation can be created in a liquid moving with a velocity $v$ less than that of a sound ($v<s$). In case of noninteracting particles the dispersion is quadratic for all $k$ and superfluidity is not possible.

\subsection{Field-theoretical approaches: $t$-matrix approximation} \label{sec_Beliaev}

To go beyond the Bogoliubov approximation, one needs to take both multiple scattering diagrams and RPA contributions into account. That can be done for example by means of a pseudopotential method \cite{Lee:1957}, or by field-theoretical methods, first applied to the Bose gas of small density at $T=0$  by \textcite{Beliaev:1958a,Beliaev:1958b} and  by \textcite{Hugenholtz:1959}.

The presence of the many particle condensate in the ground state of the interacting Bose gas was the main obstacle to application of the usual technique of Feynman diagrams to this system.
Consider for example the one-particle Green's function in the interaction representation
\begin{equation}
G(x-x')=-i\frac{<T\{\Psi(x)\Psi^+(x')S\}>}{<S>}
\label{GF}
\end{equation}
Here the average is taken over the ground state of N {\it non-interacting} Bose particles, which are all in the condensate ($N_0=N$). The $S$-matrix is expressed as usual
\begin{eqnarray}
S&=&T\Big\{ exp\Big(-\frac{i}{2}\int d^4x_1 d^4x_2 U(x_1-x_2)\times \Big. \nonumber \\
&\times  & \Big. \Psi^+(x_1)\Psi^+(x_2)\Psi(x_2)\Psi(x_1) \Big)\Big\}
\end{eqnarray}  
where $x_1$ and $x_2$ are the four-vectors, and the interaction is $U(x_1-x_2)=U(r_1-r_2)\delta(t_1-t_2)$. In order to derive the diagram series for the Green's function, we need to expand the $S$- matrix in powers of $H_{int}$. Usually the terms containing the odd number of annihilation operators vanish after averaging over the ground state, which unfortunately does not happen in the case of a Bose gas due to  the above mentioned peculiarities of the ground state. The expectation value of the $N$ product containing $a_0$ apparently does not disappear and the standard method of constructing diagrams cannot be applied in the case of an interacting Bose gas.

 This difficulty was successfully resolved by Beliaev in 1958. He noticed that for a large number of particles $N$ the diagrammatic approach can be applied to particles with momenta $p \neq 0$, while the condensed phase (which does not disappear when the interactions are turned on) can be described as a sort of external field. It is thus convenient to separate the operators $a_0$ and $a_0^+$ (which act only on the ground state) from $\Psi$ and $\Psi^+$
\begin{equation}
\Psi=\Psi'+a_0/\sqrt V; \quad \Psi^+=\Psi^{'+}+a_0^+/\sqrt V.
\label{separation}
\end{equation}
The Green's function (\ref{GF}) is then also divided into two parts, and the operations $T$ and $<...>$ are represented as two successive operations, the former acting only on $\Psi'$ and $\Psi^{'+}$, and the latter acting only on $a_0$ and $a_0^+$. The operators $a_0$ and $a_0^+$, occurring in the $S$ matrix are treated as parameters, and the expectation values over $\Psi'$, $\Psi^{'+}$ ground state can be now calculated, using standard techniques.

With these ideas in mind, Beliaev succeeded in deriving a general expression for the one-particle Green's function of the interacting system in terms of some effective self-energies $\Sigma_{ik}$ and the chemical potential $\mu$. However, 
the exact calculation of the Green's functions proved to be very complicated, and approximate methods of summing the series of Feynman graphs were developed.

For simplicity, Beliaev considered a short-range, central interaction potential $U_{\bf p}=U_0$ for $p<1/a$ and $U_{\bf p}=0$ for $p>1/a$. In the low density limit $n_0a^3\ll 1$, where $n_0$ is the density of the particles in the condensate, he obtained a crucial result, that the main contributions to the self-energies of the Green's function originate from ladder diagrams. In this case the real interaction $U$ is replaced by an effective two-particle interaction $\Gamma$, representing the sum of contributions from all ladder type Feynman graphs (Fig.3). The integral equation  for the vertex $\Gamma$, called the Bethe-Salpeter equation  is
\begin{widetext}
\begin{eqnarray}
&&\Gamma(x_1,x_2;x_3,x_4)=U_{x_1-x_2}\delta(x_1-x_3)\delta(x_2-x_4) 
\nonumber \\ &&
+ i \int d^4x_5 d^4x_6 U_{x_1-x_2}G^0(x_1-x_5) 
%\times \nonumber \\ &&\times 
G^0(x_2-x_6) \Gamma(x_5,x_6;x_3,x_4), 
\nonumber \\
\label{gamma0_space}
\end{eqnarray}
\end{widetext}
where $x\equiv({\bf r},t)$. In momentum representation this equation reads
\begin{eqnarray}
&&\Gamma(p_1,p_2;p_3,p_4)=U_{p_1-p_2}+
\nonumber \\
&&i \int d^4p_5 d^4p_6 U_{p_1-p_5}G^0(p_5)G^0(p_6) \Gamma(p_5,p_6;p_3,p_4),
\nonumber \\
\label{gamma0_mom}
\end{eqnarray}
where the momentum conservation condition $p_1+p_2=p_3+p_4=p_5+p_6$ is implied and $p_i\equiv ({\bf p_i},p_0^i)$.

\begin{figure}[!ht]
\begin{center}
\epsfxsize=0.45\textwidth{\epsfbox{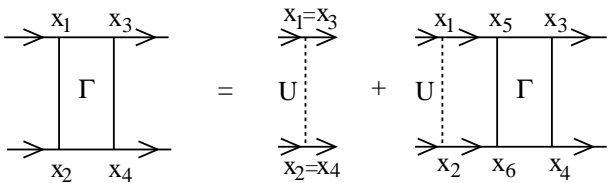}}
\end{center}
\caption{Bethe-Salpeter equation for the two-particle scattering vertex $\Gamma$.}
\label{tmatrix}
\end{figure}

It is convenient to introduce relative and total momenta according to
\begin{eqnarray}
p_1+p_2=P'; \quad p_3+p_4=P \nonumber \\
p_1-p_2=2p'; \quad p_3-p_4=2p.
\end{eqnarray}
This transformation leads to the following equation
\begin{eqnarray}
&&t(p',p,P)=U(p'-p)+ \nonumber \\
&&+i\int \frac{d^4q}{2\pi^4} U(p'-q)G^0(P/2+q)G^0(P/2-q)t(q,p,P)
\nonumber \\
\label{gamma2}
\end{eqnarray}
where we denote $\Gamma$ in the center of mass representation by ``$t$'', and the free particle Green's function is $G^0(p)=(p^0-p^2/2m+~i\delta)^{-1}$.

A conventional $t$ matrix equation is obtained from (\ref{gamma2}) after carrying out the integration over $q_0$. In two dimensions this results in the following equation
\begin{equation}
t({\bf p'},{\bf p},P)=U_{{\bf p'}-{\bf p}}-\int \frac{d {\bf q}}{(2\pi)^2}
U_{{\bf p'}-{\bf q}}\frac{t({\bf q},{\bf p},P)}{k_0^2-q^2/m+i\delta},
\label{tmatrix2}
\end{equation}
where $k_0^2=P^0-{\bf P^2}/4m$. In the scattering theory this equation is also known as the Lippmann-Schwinger equation. Physically the $t$-matrix corresponds to  the renormalization of the interaction by multiple scattering of one particle off another. 

 The standard way to treat the dilute Bose gas is thus  to replace the real potential, which is usually strongly singular, by the zero momentum $t$ matrix generated from multiple two-particle scattering, represented by the infinite summation of the ladder diagrams described above.

The $t$ matrix equation (\ref{tmatrix2}) cannot be solved explicitly, but in general its solution can be  expressed in terms of the scattering amplitude of two particles in vacuum. The scattering amplitude  $f({\bf p',p})$ for a transition from the initial relative wave vector ${\bf p}$ to a finite relative vector ${\bf p'}$ is defined by an expression
\begin{equation}
f({\bf p',p})=\int d{\bf q} U({\bf p'}-{\bf q})\Psi_{\bf p}({\bf q}),
\end{equation}
where $\Psi_p$ is a wave function of a scattering problem with potential $U$ that satisfies the following Schr\"odinger equation in momentum representation
\begin{equation}
(k^2-p^2)\Psi_k({\bf p})-\int d{\bf q} U({\bf p}-{\bf q})\Psi_k({\bf q})=0.
\end{equation}
According to  elementary scattering theory \cite{Dalfovo:1999,Castin:2001,Leggett:2001,Fetter:2002}, at low energies $s$-wave scattering becomes dominant, and the scattering amplitude $f_0$   is approximated to leading order by
\begin{equation} 
f_0\simeq \frac{4\pi \hbar^2 a_s}{m},
\label{scat}
\end{equation}
(where  the momentum dependence of the scattering amplitude can be ignored in the low energy limit). Thus at low energies, in vacuum the only remaining parameter characterizing the interaction is the $s$-wave scattering length $a_s$.

In general, the t-matrix \eqref{tmatrix2} requires the knowledge of the scattering amplitude for $k_0^2\neq q^2/m$, known as ``off-the-energy-shell'' t-matrix.  For two-particle scattering in vacuum, discussed above, only on-shell t-matrix is physically relevant. In the situation when three-body collisions become important, the calculation of the off-shell t-matrix is necessary \cite{Fadeev:1960}. In the context of the dilute Bose gases the off-shell t-matrix arises in connection with so-called many-body t-matrix approach \cite{Stoof:1993,Bijlsma:1997,Proukakis:1998}, which we discuss in the next Chapter. The many-body t-matrix takes into account the effect of the medium (mean field) in which the collisions occur. At the low energy limit the many-body t-matrix is approximated by the off-shell two-body t-matrix \cite{Morgan:2002}. The solution of the off-shell t-matrix was first proposed by \textcite{Beliaev:1958b} and \textcite{Galitskii:1958}. The alternative approach based on the inhomogeneous Schr\"odinger equation, which allows to treat the hard-sphere central potentials in one, two and three dimensions, was considered by \textcite{Morgan:2002}. \textcite{Morgan:2002} have shown for any dimension that for all potentials with a finite range, the long-wavelength limit of the off-shell t-matrix depends only on energy and not on the initial and final relative momenta of the scattered particles. This result means that low-energy collisions can be represented by a contact potential.

Consider now the  quasiparticle spectrum within the first order Beliaev approach. It turns out one can reproduce the Bogoliubov result (\ref{quasi}) with the only difference that instead of potential $U_0$ the momentum independent scattering amplitude $f_0$ appears, for in the first order $U_0$ is equal to  $f_0$.
The healing length (\ref{healing}) can then be related to a scattering length
\begin{equation}
\lambda_c=\frac{1}{\sqrt{8\pi a_sn_0}}.
\label{healing2}
\end{equation}
 The second order approximation does not modify the physical picture of the low temperature behaviour of the interacting Bose gas, but  provides the corrections to the sound velocity, and a damping proportional to $p^5$ related to the process of decay of one phonon into two. The third order corrections involve the solution of a three particle problem, which to date has not  been solved.

We now turn to the 2D system.
Following the methods developed by Beliaev, \textcite{Schick:1971} examined a two-dimensional system of hard-disk bosons of diameter $a$ at low densities and absolute zero (see also the recent study of \textcite{Ovchinnikov:1993}. The dimensionless expansion parameters are the interaction $U_0$ and the gaseous parameter $na^2$, which is small in the dilute limit. The application of Beliaev's method to 2D systems is not as straightforward as it is for 3D systems. In the 3D case, the ladder diagrams are the only contributions which do not depend on the small parameter $na^3$ and therefore it is natural to take them into account while calculating the first term in the expansions of all quantities in terms of density. In 2D the contributions from the ladder diagrams depend logarithmically on  the parameter $na^2$, in particular, the effective interaction, or $t$ matrix is proportional to $1/\ln (1/na^2)$
\begin{equation}
f_0^{2D}\sim \frac{4\pi}{m \ln(1/na^2)}.
\label{t_schick}
\end{equation}

The key conclusion of \textcite{Schick:1971} is that $1/\ln (1/na^2)$ plays the role of the small parameter in the 2D dilute system at zero temperature and the dominant contributions are derived from the diagrams of first order in this parameter. 
In this approximation he calculated the leading order correction to the chemical potential
\begin{equation}
\mu=-\frac{4\pi\hbar^2 n}{m \ln (na^2)}\left\{1+O[1/\ln (na^2)]\right\} 
\end{equation}
and the quasi-particle excitation spectrum
\begin{equation}
\xi_k=\sqrt{\mu\frac{k^2}{m}+\frac{k^4}{4m}}=\sqrt{\frac{k^4}{4m}+\frac{4\pi n}{m\ln(1/na^2)}k^2}.
\end{equation}
In the long-wavelength limit the quasi-particles behave as phonons with a speed of propagation $s=\sqrt{-4\pi n/(m\ln(na^2))}$. The spectrum changes from phonon-like to free particle-like in the vicinity of  the momentum $k_c$ defined as 
$$ka\ll k_ca\equiv -16\pi n a^2(m\ln(na^2))^{-1} \ll 1 \; . $$

The ground state energy per particle and the condensate fraction take the form \cite{Schick:1971}
\begin{eqnarray}
E/N&=&-\frac{2\pi\hbar^2 n}{m \ln (na^2)}\left\{1+O[1/\ln (na^2)]\right\} \cr
\frac{n_0}{n}&=&1+\frac{1}{\ln (na^2)}+O\left[1/(\ln (na^2))^2\right].
\end{eqnarray}

\subsection{Gross-Pitaevskii mean-field theory} \label{sec_GP}

The ground state and thermodynamic properties of an interacting Bose system  confined to an external potential $V_{ext}=\frac{1}{2}\hbar\omega_{h0}(r/a_{h0})^2$ ($a_{h0}$ is the trap size (\ref{trapsize})) can be directly calculated from the Hamiltonian
\begin{eqnarray}
&&H=\int dr \psi^+(r)\left(-\frac{\hbar^2}{2m}\bigtriangledown^2+V_{ext}(r) \right)\psi(r) +
\nonumber \\
&&+\frac{1}{2}\int dr dr'\psi^+(r)\psi^+(r')U(r-r')\psi(r')\psi(r) \qquad
\end{eqnarray}
using numerical methods, such as  quantum Monte Carlo. 
Nevertheless for most experimentally relevant situations (when the number of atoms is large) the mean-field description of the system proves to be sufficient. 
 In this case the macroscopic low energy behaviour of the system can be explored under the assumption that the order parameter varies over distances larger than the mean interparticle spacing.

Such a mean-field approximation was first developed by Gross and Pitaevskii. Their approach, which is valid  
 in the dilute limit, is a straightforward generalization of Bogoliubov theory for the gas in the trap.
One should bear in mind that the diluteness condition $n_{max}a_s^3 \ll 1$ does not automatically secure the weakness of the interactions. The interaction strength is specified by an extra parameter (see in particular the review of \textcite{Dalfovo:1999} and the paper by  \textcite{Fetter:1999}). The interaction energy, which is of the order of $gNn$ is to be compared with the kinetic energy, proportional to $Na_{h0}^{-2}$. Since the average density of atoms $n \sim N/a_{h0}^3$, the interaction strength can be characterized by a dimensionless parameter $N|a_s|/ a_{h0}$. 
When $Na_s\ll a_{h0}$, it means that the coherence length $\lambda_c$ (\ref{healing}) is large in comparison with the size of the trap $a_{h0}$ and the system is assumed to be nearly ideal and is described by a Gaussian distribution (\ref{spwf}). In the opposite limit, $Na_s\gg a_{h0}$, the coherence length is small and the dilute gas exhibits important nonideal behaviour \cite{Dalfovo:1999}. 

The mean-field Gross-Pitaevskii approximation is extensively presented  in the literature (see for instance review by  \textcite{Dalfovo:1999}, and paper by \textcite{Leggett:2003}, and a review with an emphasis on experiment by \textcite{Angilella:2006}), therefore we only  mention briefly the key concepts of its derivation.  
Gross and Pitaevskii's approach is based on the Bogoliubov prescription for the condensate (\ref{prescription}), according to which 
 the boson field operators $\psi$ are written as a sum of a classical field $\phi$, having the meaning of the order parameter, and a small perturbation $\psi'$
\begin{equation}
\psi(r,t)=\phi(r,t)+\psi'(r,t),
\label{wfGP}
\end{equation}
implying that the depletion of the condensate is small. As a side note, we mention that in principle  the problem of the order parameter definition in a finite inhomogeneous system arises in this case, but it turns out that the wave function of the condensate has a clear meaning, if determined through the diagonalization of the one-body density matrix in analogy with liquid-helium drops \cite{Lewart:1988}. This issue is also discussed in detail in a review by \textcite{Leggett:2001}.

One can expand the theory in the parameter $\psi'$ and derive the  equation for $\phi$ either from the standard Heisenberg equation, or alternatively by taking the variation of the classical action $S$ of the type $S=\int dt d{\bf r} \overline \phi \left[i\partial_t -\frac{\hbar^2}{2m}\nabla^2-V_{ext}-\frac{g}{2}\overline \phi \phi\right]\phi $ 
 with respect to $\overline \phi$ (saddle point approximation). The derivation of the Gross-Pitaevskii (GP) equation and the next order corrections within the bosonic field theory can be found in the paper by \textcite{Stenholm:1998}. 

The resulting Gross-Pitaevskii equation is 
\begin{equation}
i\hbar\frac{\partial}{\partial t}\phi(r,t)=\left(-\frac{\hbar^2}{2m}\bigtriangledown^2+V_{ext}+g|\phi(r,t)|^2 \right)\phi(r,t),
\label{GPeq}
\end{equation}
where we have approximated the potential by a $\delta$-function, $V(r-r')=g\delta(r-r')$ (which we can do under the assumption that the interparticle spacing is much larger than the interaction range), and where $g=\frac{4\pi\hbar^2a_s}{m}$ is the 3D coupling constant. 
This coupling constant is equal to the zero momentum limit of the scattering amplitude  (\ref{scat}) discussed above.

In the limit $N\rightarrow \infty$ (Thomas-Fermi approximation) the kinetic energy contribution can be neglected and the Gross-Pitaevskii equation can be solved analytically. This classical Thomas Fermi approximation breaks down in the vicinity of the boundary of the condensate, where the gradient of the condensate density is no longer small.

We discuss now the coupling constant of the 2D Bose gas. It was first demonstrated by Y. Lozovik in 1971 (see the review of \textcite{Petrov:2004}) that to zero order in perturbation theory the coupling constant $g^{2D}=\hbar^2 f_0^{2D}/m$, where $f_0^{2D}$ is the scattering amplitude at energy of the relative motion $E=2\mu$. 

This coupling constant can be treated as a parameter, as in the work by \textcite{Bayindir:1998} (see also references therein), where the two-dimensional Bose gases described by the GP equation have been studied. For some range of interaction strength it was shown that interacting bosons behave similarly to the noninteracting case in a harmonic trap. For  weak short-range interparticle interactions, a finite temperature BEC phase transition was found to occur. 

On the other hand, the coupling constant in 2D is expected to display a logarithmic dependence on density (cf.(\ref{t_schick})) in accordance with estimations by \textcite{Schick:1971} for $f_0$ in case of a homogeneous gas. 
The precise choice of $g^{2D}$ has in fact been a controversial issue (see \textcite{Lieb:2001} and references therein). For example, \textcite{Kim:1999}
suggested $g^{2D}\sim 1/\ln(1/ka)$, where $0< ka\ll 1$ and $k$ is the infrared cut off introduced by the trap at $1/a_{h0}$, so that  $g^{2D}\sim 1/\ln(a_{h0}/a)$. This kind of approximation may be reasonable when the size of the trap is much larger than all other length scales in the problem. 

Note, that for quasi-2D gas in a trap the coupling constant was derived by \textcite{Petrov:2000}
\begin{equation}
g^{Q2D}=\frac{2\sqrt{2\pi}\hbar^2}{m}\frac{1}{a_{ho}/a+(1/\sqrt{2\pi})\ln(1/\pi k^2a_{h0}^2)}.
\end{equation}

The rigorous derivation of the Gross-Pitaevskii functional for a two-dimensional interacting gas was provided by \textcite{Lieb:2001}. Their analysis leads to the following expression for the coupling constant
\begin{equation}
g^{2D}=\frac{1}{|ln(\bar n a^2)|},
\label{g2d_lieb}
\end{equation}
where $\bar n$ is the average density of the particles, proportional to $\sqrt{N}$. The mean density is defined as $\bar n=1/N\int n^{TF}(r)^2d^2r$, with the Thomas-Fermi density being $n^{TF}(r)=(\mu^{TF}-V_{ext}(r))/8\pi$, and $\mu^{TF}$ chosen so that the constraint $\int n^{TF}=N$ holds. The density expansion has been applied to the case of a  2D Bose gas at zero temperature by \textcite{Cherny:2001} in order to derive the Gross-Pitaevskii equation.

The modification of the GP equation due to the many-body renormalization of the scattering, mentioned in the section \ref{sec_Beliaev}, has been provided by \textcite{Lee:2002}. The effective interparticle interaction in 2D is modeled by the off-shell two-body t-matrix, that at low energies depends on the energy of the collision. The energy dependence of the effective interaction can be written in the density-dependent form and applied to trapped 2D gas. This leads to the GP equation, describing the condensate wave-function that no longer has a cubic non-linearity in $\Psi$, but instead goes as $(|\psi|^2/\ln|\psi|^2)\psi$ \cite{Lee:2002}.

It is also interesting to analyze the deviations from the  mean-field behaviour, since the experimental system is well controlled nowadays and different regimes can be realized. The corrections to the mean-field ground state solution stem from the quantum fluctuations, and their effect becomes more prominent with the growth of the gas parameter, as has been observed in Monte Carlo simulations.  For the calculation of quantum corrections in a systematic way we refer the reader to the paper by \textcite{Andersen:2002a} and references therein. Many references on the GP approximation and beyond can be found elsewhere \cite{Angilella:2004c,Kolomeisky:2000}. For the effects of a third spatial dimension and the self-consistent calculation of the coupling constant see the paper of \textcite{Cherny:2004} and references therein.

\section{Finite temperature problems} \label{Finite}

Zero-temperature techniques are not really suitable for controlling IR thermal fluctuations, and new methods have to be devised to describe the interacting system at finite $T$. 
At the beginning of the Chapter \ref{Finite} we present the generic properties of the 2D XY models and the concept of quasi-long-range order, which is the central concept in the phase transition theory in 2D. Why the true long-range order cannot form in  2D uniform system, we discuss in detail in section \ref{LRO2D}, and especially the way familiar concepts from 2D phase transition theory should be revised in the trapped case. 

 In Section \ref{sec_Popov} we present the theory of Popov, who pioneered the finite-temperature generalization of Beliaev's field-theoretic approach, and described the low-temperature superfluid state of the 2D Bose gas. In Section \ref{sec_dilute} we show how the diluteness condition of Fisher and Hohenberg, discussed in the introduction, arises as an applicability limit of the Popov's t-matrix approach. In Section \ref{sec_MBT} we describe methods, which generalized and/or improve the results of Popov, and also the Monte Carlo simulations, which are up to date most reliable numerical calculations of the superfluid phase of the 2D Bose system. Before concluding, we mention how unique the 2D system with a contact interaction is, for it possesses an inherent symmetry, which leads to the birth of the special breathing modes, which in principle can be checked experimentally.

\subsection{Introduction: 2D XY models} \label{intro_XY}

For our further analysis it is important to recognize that a uniform, interacting Bose system belongs to the XY universality class, characterized by a vector order parameter  (for a comprehensive analysis see the book by \textcite{Chaikin:1995}). 
It means that the finite-temperature behaviour of the 2D Bose gas is determined by generic properties of the 2D XY model.

We know, that the 2D XY models are very special, for the long-range thermal fluctuations destroy the long-range order at finite temperatures (Bose-Einstein condensation in case of 2D Bose gas). The existence of these long-wavelength modes in a 2D Bose fluid was first pointed out by Bogoliubov in his ``$k^{-2}$ theorem'' in 1961, and later confirmed by \textcite{Hohenberg:1967} and by \textcite{Mermin:1966} (this issue is discussed in more detail in section \ref{LRO2D}). 

However, a special type of order - topological order - which gives rise to superfluidity, can develop in a two-dimensional Bose fluid below the ``Kosterlitz-Thouless''  temperature $T_{KT}$, as predicted by \textcite{Kosterlitz:1973} and \textcite{Berezinskii:1970,Berezinskii:1971} using the renormalization group method (RG). Below $T_{KT}$ the continuous U(1) symmetry (rotations in a two-dimensional plane) is broken and the system acquires a finite rigidity, or phase stiffness $\rho_s$. The order parameter correlations decay algebraically (for any coupling of the XY model), and the average order parameter is zero. However locally, the order parameter can have a well-defined value. This unique situation is described in terms of quasi-long-range order (QLRO) \cite{Chaikin:1995}. Important low-lying excitations of the QLRO phase are vortex pairs (two vortices with opposite winding numbers) whose fugacity decreases with distance, thus not destroying the connectivity of the state (therefore $\rho_s\neq 0$). 

The phase transition to a disordered state (with $\rho_s=0$) is associated with a dissociation of the coupled vortex pairs. Above $T_{KT}$ the vortex fluid can be treated as a kind of vortex plasma, where vortices play the role of mobile ``charges'', interacting via a Coulomb potential. In this language the state below $T_{KT}$ can be described as an ``insulating'' state of bound ``charges''. The mapping of the 2D XY model onto the two-dimensional Coulomb gas is considered in detail in the review by \textcite{Minnhagen:1987}.

The rigidity, or superfluid density $\rho_s$ does not go continuously to zero at the critical temperature, but experiences a universal jump 
\begin{equation}
\frac{m^2k_BT_{KT}}{\hbar^2\rho_s(T_{KT_-})}=\frac{\pi}{2},
\label{uni_jump}
\end{equation}
first predicted by \textcite{Nelson:1977} and successfully verified in experiments on superfluid $^4$He films, absorbed on a substrate \cite{Bishop:1978}. 

An interesting interpretation of Kosterlitz-Thouless physics in the context of bosonic systems was put forward by \textcite{Kagan:1987} almost 20 years ago. They propose that below $T_{KT}$ the 
system forms a ``quasi-condensate'', a condensed state achieved in a local sense. The introduction of the quasi-condensate concept was motivated by a peculiar behaviour of the one-particle density matrix $\rho(r)$ at large distances in 2D (Fig.4).

\begin{figure}[!ht]
\begin{center}
\epsfxsize=0.45\textwidth{\epsfbox{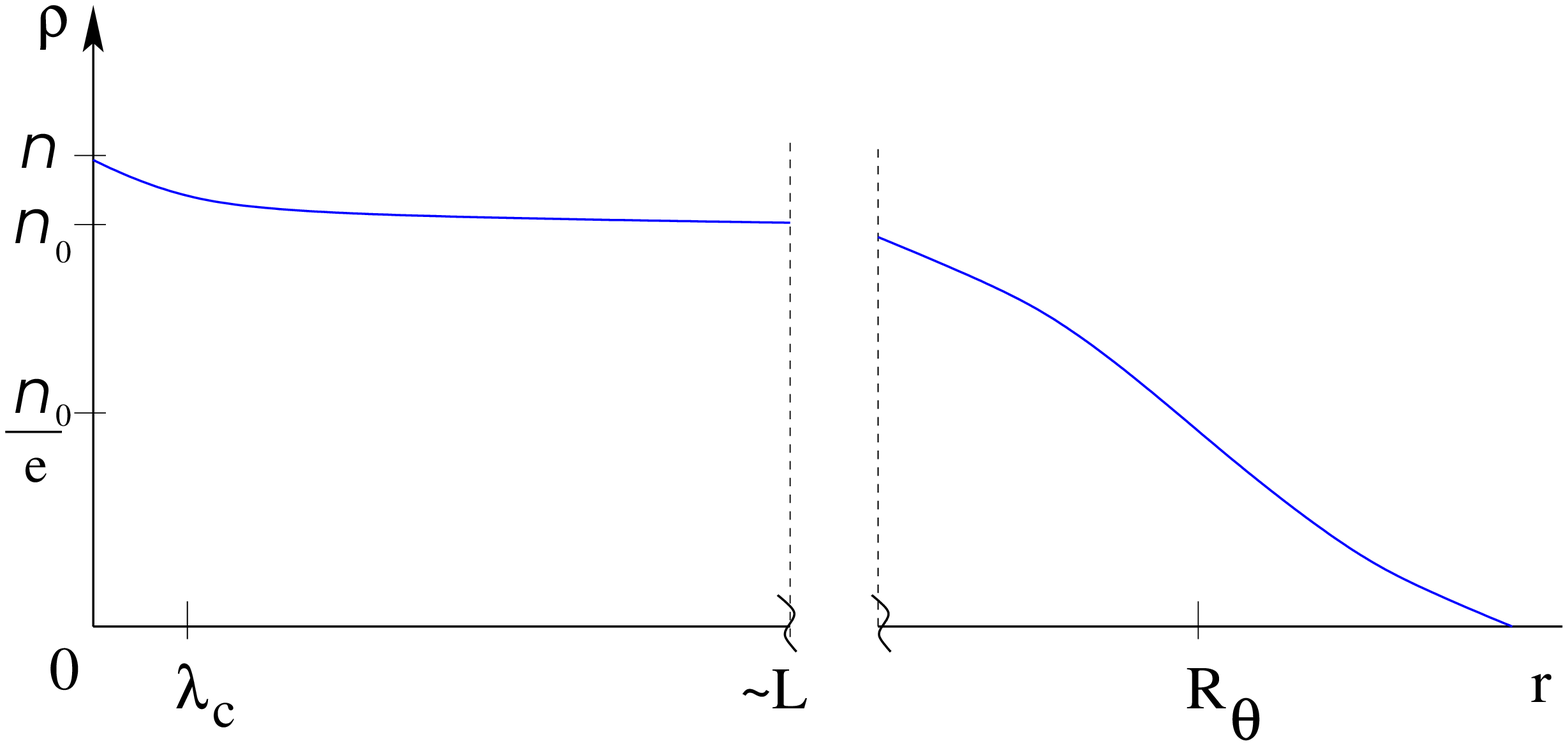}}
\end{center}
\caption{One-particle density matrix $\rho(r)=\langle \psi^+(0)\psi(r)\rangle $ in two dimensions. Two characteristic length scales $R_{\theta}$ and $ \lambda_c$ are shown, $R_{\theta}\gg \lambda_c$. At large distances $\lambda_c\ll L \ll R_{\theta}$ the one-particle density matrix is equal to the condensate density $n_0$.}  
\label{tmatrix}
\end{figure}

There are two length scales associated with the behaviour of $\rho(r)$: the aforementioned correlation length $\lambda_c$ at which $\rho(r)$ relaxes from the value $n$ at $r=0$ to $n_0$, and the characteristic radius of the phase fluctuations $R_{\theta}$, which is rather large $R_{\theta}\gg\lambda_c$. The appearance of large $R_{\theta}$ can be understood in the following way: at large distances $\rho$ falls off as a power law of $r$ (\textcite{Kane:1967}) $\rho(r)\sim n_0(r/r^*)^{-\alpha}$, where $r\gg r^*$ and the coefficient $\alpha$ is proportional to the temperature and the Schick's parameter $\alpha\sim T/(T^*ln(1/na^2))$ ($T^*\sim (2m\lambda_c^2)^{-1}$) and therefore is very small $\alpha\ll 1$. 
As a result of this the density matrix $\rho$ decays over a large length scale $R_{\theta}\sim r^*e^{1/\alpha}$ \cite{Kagan:1987}.

Conceptually, the system can be divided into blocks of size $L$, which is smaller than $R_{\theta}$.
In each block one can introduce the  wave-function of the condensate with a well-defined phase.
The whole system is then described in terms of an ensemble of wave-functions of the blocks.  
Condensate wave-functions within the ensemble corresponding to blocks separated by a distance greater than $R_{\theta}$ have uncorrelated phases, and
  it is impossible to define the condensate wave function for the entire system as a whole. The state of matter  with a fluctuating phase is called  a ``quasi-condensate'' \cite{Kagan:1987}. See also the extention of Bogoliubov methods to quasi-condensates by \textcite{Mora:2003}. 

What happens to the XY universality class concepts in an experimentally realizable system of cold atoms confined in a trap, remains a controversial issue. We will see in the next Section, that many issues should be crucially reformulated in order to address the physics of trapped cold gases.

\subsection{Problems  of the long-range order formation in 2D} \label{LRO2D}

The notion that the development of the long-range order (LRO) is not possible in 2D  dates back to the work of \textcite{Peierls:1935}, who argued that the thermal motion of low energy  phonons will ruin the LRO in a 2D solid. A rigorous proof of the Peierls' statement was provided later by \textcite{Mermin:1968}.

Subsequent work by \textcite{Mermin:1966} provided a proof that there is  no spontaneous magnetization or sublattice magnetization in an isotropic Heisenberg model with finite range interactions. At the same time \textcite{Hohenberg:1967}  succeeded in ruling out the existence of a  conventional superfluid or superconducting ordering in one and two dimensions. It was also shown by \textcite{Coleman:1973}, that ``there are no Goldstone bosons in 2D'', which is equivalent to saying that there is no LRO in 2D.

A rigorous proof of the Mermin-Wagner-Hohenberg results exploits the  Bogoliubov and Schwartz inequalities (Appendix I) and leads to the following result for the average occupation number of ${\bf k}$ states
\begin{equation}
\langle a_k^+a_k\rangle\equiv n_k\geq -\frac{1}{2}+\frac{mTn_0}{k^2n}.
\label{ineq}
\end{equation}
Here $n_0$ is a condensate density and $n$ is a total density. It is clear now that the appearance of the condensate (macroscopic occupation of a single state) in 2D for finite temperatures fails due to the mathematical fact that the function $k^{-2}$ is not integrable at small momenta in two-dimensional $k$ space. Physically, the long-range thermal fluctuations prevent the formation of a coherent condensate. 

The same result can be obtained from the infrared asymptote of the one-particle Green's function at zero frequency
\begin{equation}
G({\bf k},0)\approx -\frac{n_0m}{n_sk^2},
\label{asymp}
\end{equation}
and was first derived by \textcite{Bogoliubov:1961}. The derivation of the asymptotic behaviour (\ref{asymp}) in a functional integrals approach can be found in \textcite{Popov:1983}. Since the Green's function defines the average number of particles with momentum $k$, it is readily seen we arrive to the same result (\ref{ineq}). 
The statement, that the condensate does not appear in a 2D interacting Bose system at any finite temperature, is also known as the {\it Bogoliubov} $k^{-2}$ {\it theorem}. 

We have already mentioned that in the context of modern condensed matter theory the absence of the LRO in 2D is discussed in terms of general properties of the XY models. A respective direction or a phase of the $d$-dimensional XY  order parameter is specified by an angle $\theta$.  The variance in the fluctuation of the order parameter phase  is given by the integral
\begin{equation}
\langle \theta^2({\bf r}) \rangle \sim \frac{T}{\rho_s} \int \frac{d^dq}{(2\pi)^d}q^{-2}=\frac{T\Lambda^{d-2}}{\rho_s(d-2)},
\end{equation}
where $\Lambda$ is the wave number cutoff \cite{Chaikin:1995}. It is readily seen that $d=2$ is the critical dimension of the XY universality class and the fluctuations destroy long-range order in the 2D $XY$ model in accordance with the conclusions of Bogoliubov, Mermin, Wagner and Hohenberg. Quasi-long-range order, discussed in the previous section is nevertheless possible in 2D.   

In case of a trapped gas the Bogoliubov-Mermin-Wagner-Hohenberg (BMWH) theorem rules out BEC in 2D in the interacting system (see \textcite{Mullin:1997}). However, the question arises if one can actually apply BMHW theorem to a system confined within a harmonic potential. Is it still possible to unambiguously rule out the condensate formation in 2D atom traps? The applicability of the BMWH theorem to the inhomogeneous case requires careful consideration, for  the  Bogoliubov-Hohenberg inequality was derived assuming an infinite uniform system. In this approximation, many special features of practically realized condensates, such as their formation in real space, are excluded.

An alternative version of the Hohenberg inequality, suitable for the experimentally realizable Bose systems, has been recently proposed by \textcite{Fischer:2002,Fischer:2005}. Taking the dimension of the trap to be an experimentally controlled parameter, Fischer addressed the issue of a spatially localized Bose condensate, with the question in mind of how far one could ``stretch'' the 3D condensate cloud before the coherence will be destroyed. Fischer derived an inequality, which controls the size of the smallest possible condensate for a given condensate and density profile. In Appendix I we briefly sketch the underlying concepts of his derivation. 

The resulting inequality reads
\begin{equation}
\frac{n-n_0}{n_0}\geq \frac{2\pi R_c^2}{n\lambda_{dB}^2}C({\bf k})-\frac{1}{2n_0}\left(1-|\psi_0({\bf k})|^2/V_0) \right),
\label{Uwe_in}
\end{equation}
where $\psi_0({\bf k})$ is the condensate wave-function, $R_c$ is the effective radius of the condensate wave function (effective radius of the curvature of the condensate)
\begin{equation}
R_c=\left(V_0/n\int d^dr\psi_0(r)[-\Delta_r\psi_0^*(r)]n(r) \right)^{-1/2}
\end{equation}
and
\begin{equation}
C({\bf k})=\left|\int d^dr|\psi_0|^2exp(ikr)-\psi_0(k)\int d^dr\psi_0^*(r)|\psi_0(r)|^2\right|^2.
\end{equation}
Note, that  only requirement on the Hamiltonian of the system, that is needed to derive the inequality (\ref{Uwe_in}) is that it should not  contain any explicit velocity dependence in the interaction and external potentials.

Since in 2D $R_c^2$ scales as $n$, this case can be considered as marginal and the condensate still can emerge even in an interacting system. This is because the usual log divergences inherent for 2D are cut off by a trap.  The inequality (\ref{Uwe_in}) is a geometrical equivalent of the Bogoliubov-Hohenberg inequality, since it gives the lower bound for the ratio of the effective radius of the condensate to the de Broglie wavelength $\lambda_{dB}$. The second term of rhs (\ref{Uwe_in}) can be used to obtain an upper limit on the possible condensate fraction as a function of temperature. Concrete examples of the application of (\ref{Uwe_in}) to quasi-1D systems are given in the paper by \textcite{Fischer:2002}.

One can also approach the problem of the condensate formation by directly analyzing  the phase fluctuations of the order parameter (for a review see \textcite{Hellweg:2001}). Phase fluctuations are caused by the thermal excitations and are  always present at finite temperatures. Note, that at very low temperatures density fluctuations in equilibrium are suppressed due to their  energetic cost and can therefore be ignored. This assumption is not valid in the vicinity of a vortex core, but at very low temperatures, the vortex formation is negligible.

As an aside, we mention that the concept of phase in quantum systems, introduced by Dirac as a canonical conjugate observable to the number operator $\hat n$, remains a controversial issue in certain circles. Formally it is known, that if $\hat n$ is an operator with a purely discrete spectrum (which is always true for the number operator), then there can exist no operator $\hat \theta$ such as the commutator $[\hat n, \hat \theta]=i\hat 1$ holds. Different versions of phase-related operators  have been  constructed in order to overcome this difficulty (see for example the review by \textcite{Carruthers:1968} and the textbook on Quantum optics by \textcite{Mandel:1995}). Alternatives to conventional symmetry breaking approaches have even been proposed (see the paper of \textcite{Stenholm:2002} and references therein ). An intriguing suggestion that the interference patterns of two atomic condensates can be  explained without ever evoking the notion of phase was put forward by \textcite{Javanainen:1996}. 

In the present article we adopt the ``conventional'' and certainly more convenient approach, according to which the bosonic field operator takes on the form 
\begin{equation}
\psi({\bf r})=\sqrt{n_0({\bf r})}\exp[i\theta({\bf r})]
\label{BECOP}
\end{equation}
for the large number of particles. Here $\theta({\bf r})$ is the operator of the phase fluctuations and $n_0({\bf r})$ is the condensate density at $T=0$. 

To proceed with calculations it is convenient to expand the phase operator in terms of the creation and annihilation operators for Bogoliubov quasiparticles (see \textcite{Shevchenko:1992})
\begin{equation}
\hat \theta({\bf r})=\frac{1}{2\sqrt{n_0({\bf r})}}\sum_k\left( (u_k+v_k)\hat a_k+(u_k-v_k)\hat a_k^{+}\right), 
\label{phaseexp}
\end{equation}  
where $a_k$ is the annihilation operator for the Bogoliubov excitation with energy $\epsilon_k$, and $u_k$, $v_k$ are excitation functions, determined by a bosonic equivalent of the Bogoliubov-de Gennes equations (for a general reference see the book of \textcite{de Gennes:1966} ). Expression (\ref{phaseexp}) can be obtained in the formalism of Bogoliubov transformation generalized to an inhomogeneous case. 

Phase fluctuations  in a quasi 2D system can be analyzed  within the formalism of the one-particle density matrix (see the works by \textcite{Petrov:2000,Petrov:2001})
\begin{equation}
\langle \psi^+({\bf r})\psi(0) \rangle=\sqrt{n_0({\bf r})n_0(0)}\exp[-\langle(\Delta \theta({\bf r}))^2 \rangle/2)]
\end{equation}
One should mention that the quasi two-dimensionality of the system implies that the scattering of particles acquires a 3D character, while the kinetic properties of the gas remain two-dimensional.

It is clear from (\ref{phaseexp}) that the estimation of the phase fluctuations $\langle \Delta \theta({\bf r})^2 \rangle$ requires a knowledge of the Bogoliubov quasiparticle spectrum  in the inhomogeneous systems (see papers of \textcite{Stringari:1996} and  \textcite{Oenberg:1997} and references in papers by \textcite{Petrov:2000,Petrov:2001}). This spectrum  is discrete for $T\ll \mu$ and for $T\gg \mu$ one can use the local density approximation. In the Thomas-Fermi regime for $T\gg \mu$ one obtains the following approximation 
\begin{equation}
\langle \Delta \theta({\bf r})^2 \rangle \sim T \ln(R/\lambda_{dB}).
\label{condphase}
\end{equation}
Note, that (\ref{condphase}) does not depend on precise expression for the repulsive coupling constant. 

From (\ref{condphase}) one can estimate  the characteristic radius $R_{\theta}$ of phase fluctuations (the characteristic length at which phase changes by $2\pi$) to be $R_{\theta}\simeq \lambda_{dB}exp(T_{\theta}/T)$ with  $\quad k_BT_{\theta}=N(\hbar\omega_{\perp})^2/\mu$. We thus arrive at the conclusion that at low temperatures $T\ll T_{\theta}$ the characteristic radius of the phase fluctuations is larger than the size of the trap $R_{\theta}\gg R_{\perp}$, so a true condensate exists. 
The emergence of a true condensate is attributed to the weakening of phase fluctuations induced by a trap, which introduces a low momenta cut off into the excitations in the system.
At higher temperatures, $T\gg T_{\theta}$  the system is characterized as a quasicondensate ($R_{\theta}\ll R_{\perp}$). 

The crucial effect of a trap for 2D Bose gases was also emphasized by \textcite{Ho:1999}.  They pointed out that 
long wavelength quantum fluctuations will be partially suppressed due to the gapped spectrum of collective modes \cite{Stringari:1996} and off-diagonal order will survive in 2D.

 Since there is an experimental evidence in support of BEC existence in 2D, the discussion is not yet closed. 
Quantum Monte Carlo simulations for bosons in a two-dimensional harmonic trap do indeed show that a significant fraction of the particles is still present in the lowest state at low energies \cite{Heinrichs:1998}.

\subsection{Popov's approach} \label{sec_Popov}

In this section we consider the way \textcite{Popov:1983} generalized the field-theoretical methods, developed by Beliaev to finite temperatures. It is curious that the method, suggested by Popov in 1965, is conceptually very similar to renormalization-group approach, successfully applied in the 1970th to phenomena, unaccessible to perturbative methods, such as Kondo effect \cite{Hewson:1993}.   

As usual one starts with the introduction of the temperature Green's function 
\begin{eqnarray}
G(x,\tau ; x' ,\tau ') &=&-<\psi(x,\tau)\overline \psi (x',\tau')> \cr
&=&-\frac{\int e^S\psi(x,\tau)\overline \psi (x',\tau')d\psi d\overline \psi}{\int e^S d\psi d\overline \psi},
\label{Grfunc} \qquad
\end{eqnarray}
where  $S$ is the classical action of the Bose gas
\begin{equation}
S=\int_0^{\beta}d\tau \int  d^3x \overline \psi(x,\tau) \partial_{\tau} \psi(x,\tau)-\int_0^{\beta}d\tau H(\tau)
\label{action}
\end{equation}
and 
\begin{eqnarray}
&&H(\tau)=\int d^3 x  \overline \psi(x,\tau)(-\frac{\nabla^2}{2m}-\mu) \psi(x,\tau)+ \cr
&&+\frac{1}{2}\int d^3x d^3y U(x-y) \overline \psi(x,\tau) \overline \psi(y,\tau) \psi (y,\tau) \psi(x,\tau).
\nonumber \\
\end{eqnarray}

Next step is the construction of the perturbation theory and corresponding diagrams arising from integrals of the type (\ref{Grfunc}), by performing the usual ``trick'' of separating out the condensate operators (\ref{separation}). However, in case of Bose system the perturbation series converges very poorly for small momenta and frequencies. In other words, the infrared asymptote of the Green's function is singular. In order to avoid these difficulties, Popov suggested following modifications: the bosonic fields $\psi$ 
\begin{equation}
\psi(x,\tau)=\sqrt{\frac{1}{\beta V}}\sum_{k,\omega}exp[-i(kx-\omega\tau)]a(k,\omega)
\end{equation}
is divided into a short wavelength ``fast'' component $\psi_1$ and a long wavelength ``slow'' component $\psi_0$ ($\psi(x,\tau)=\psi_0(x,\tau)+\psi_1(x,\tau)$) (see Fig.5).
The momentum $k_0$ which separates the slow modes from the rapidly oscillating modes depends on the particular Bose system and only its order of magnitude can be estimated. Introduction of $k_0$ removes the divergences at small momenta, regularizing the perturbation theory. 

\begin{figure}[!t]
\begin{center}
\epsfxsize=0.45\textwidth{\epsfbox{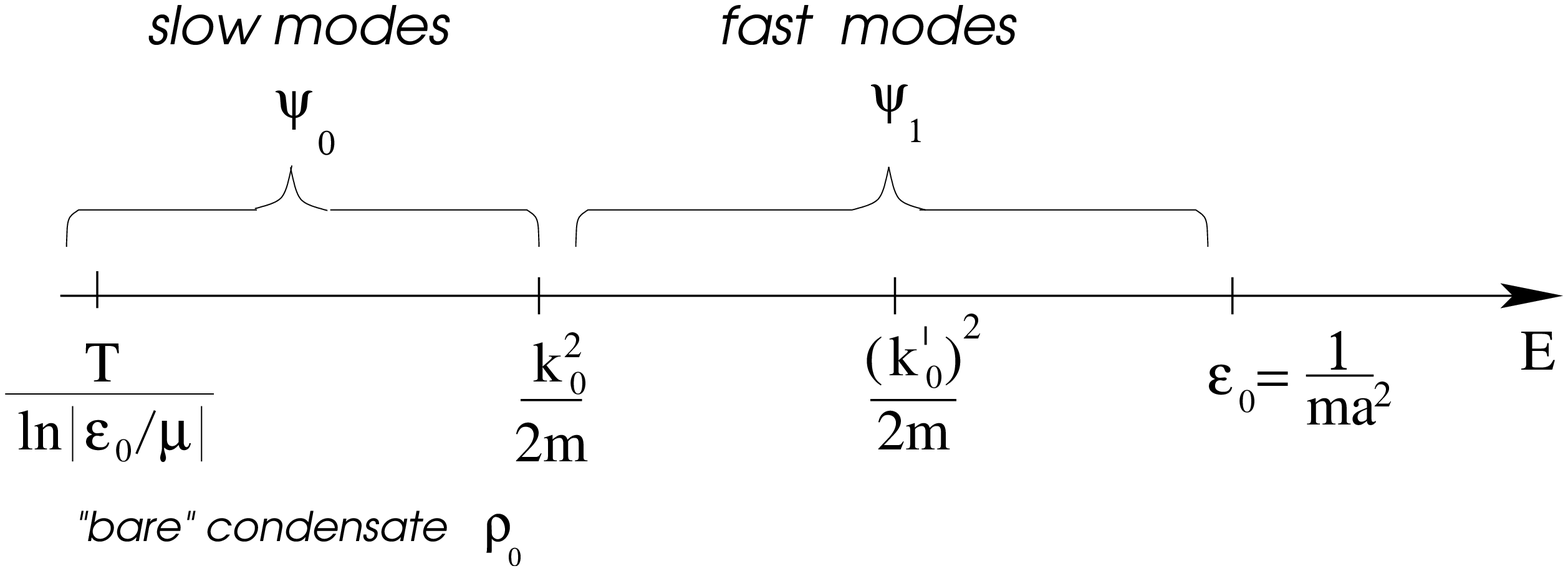}}
\caption{Energy scales in Popov's approach: $T/\ln|\epsilon_0/\mu|\ll k_0^2/2m\sim T \ll (k_0^{'})^2/2m\ll \epsilon_0=1/ma^2$.}
\label{popovfig}
\end{center}
\end{figure}
A method of successive integration, first over the ``rapid'' and then  over the ``slow'' fields, is then applied, using different schemes of perturbation theory at different stages of the integration 
(see chapter 4 in \textcite{Popov:1983}). The fast modes ``see'' the slow modes as an effective condensate (``bare'' condensate according to Popov) with a superfluid density $\rho_0=|\psi_0|^2$. 
In Appendix II we give a succinct derivation of main Popov's results.

This method of subsequent integration, developed by Popov, allows to  estimate the low temperature asymptotic behaviour of the one -particle Green's function, and to derive a power-law decay of $G(x,y)\sim|x-y|^{-\alpha}$ for $|x-y|\rightarrow \infty$ (in 1D and 2D)  rather than the exponential decay that occurs at high temperatures. In 2D, as we have mentioned in \ref{intro_XY} this  signals the development of  topological LRO at low temperatures. 

The analysis is based on the t-matrix description of the effective interactions, and the key property of the 2D t-matrix is that at low energies it vanishes, and at high energy cut-off the t-matrix diverges (see Appendix II).  This  results in an extremely small critical temperature
\begin{equation}
T_c\simeq \frac{\mu \ln(\epsilon_0/\mu)}{4 \ln \ln(\epsilon_0/\mu)},
\end{equation}
where $\epsilon$ is a high energy cut-off and $\mu$ is a chemical potential. 
Bear in mind, that this is a mean-field derivation and the condition for the superfluid transition was assumed to be $\rho=\rho_n$, because Popov (as well as \textcite{Berezinskii:1970,Berezinskii:1971}) thought  that at the critical temperature $T_c$, the superfluid density vanishes. 

The applicability of Popov's mean-field description is based on the assumption of a very small exponent $\alpha$. For large $\alpha$ the probability of creation of quantum vortices becomes big and even this modified perturbation theory is invalid (see also discussion in \ref{sec_dilute} and the ``corrected'' many-body mean-field theory in \ref{sec_MBT}). The applicability of the Popov's t-matrix description and the diluteness condition, derived by Fisher and Hohenberg is the main subject of Section \ref{sec_dilute}.

\subsection{Diluteness condition and validity of $t$-matrix approximation} \label{sec_dilute}

We have already discussed that the perturbative treatment of the dilute weakly interacting Bose gas amounts to replacing the real potential by an effective two-particle $t$-matrix, obtained by summing up  all ladder diagrams. From this point of view the diluteness condition determines the range of validity of the $t$-matrix approximation. 

An explicit form for the diluteness condition of 2D interacting Bose gas at finite temperatures was first introduced by \textcite{Fisher:1988}. They pointed out that singularities inherent to 2D systems (vanishing of scattering $t$ matrix at zero temperature and classical divergence of phase fluctuations) might lead to drastic modifications of the usual dilute gas expansion.

As we have discussed (see section \ref{sec_Beliaev}) at zero temperature in 2D the diluteness condition $na^2\ll 1$ is replaced by 
\begin{equation} 
\frac{1}{\ln (1/na^2)} \ll 1.
\label{dilute_s}
\end{equation}

Popov's theory can be used to demonstrate that at finite temperature, the above condition (\ref{dilute_s}) is replaced by an even more stringent inequality \cite{Fisher:1988}
\begin{equation}
\ln \ln\frac{1}{na^2}\gg 1.
\label{dilute_FH}
\end{equation}

Fisher and Hohenberg (FH) provided a heuristic derivation of this result, based on the Bogoliubov quasiparticle picture. Their analysis is based on the simple observation that the usual Landau quasiparticle formula for the superfluid density
\begin{equation}
\frac{\rho_s}{\rho}=1-\frac{\beta}{\rho d} \int \frac{d^dk}{(2\pi)^2}k^2\frac{e^{\beta \xi_k}}{(e^{\beta \xi_k}-1)^2},
\label{Landau}
\end{equation}
where $d$ is the dimension, does not have any singularities for $d=2$, except in case when the chemical potential is small ($\mu$ is introduced in (\ref{Landau}) via the Bogoliubov quasi-particle spectrum $\xi_k^2=n_0U_0k^2/m+k^4/4m^2\equiv \mu k^2/m+k^4/4m^2$. The validity of this approximation for $\mu$ is discussed in \cite{Beliaev:1958a,Beliaev:1958b}). By introducing the infrared cut off ($k_0\sim \sqrt{\mu}$) via the ansatz
\begin{equation}
\mu\sim \frac{n}{|ln(a^2 \mu)|},
\end{equation}
the regularization of the integral of Eq.(\ref{Landau}) can be achieved, and one arrives at Popov's equations for the ``superfluid'' and ``normal'' densities (\ref{totaldensity})-(\ref{densities}).

The analysis of the temperature dependence of the superfluid density allows one to separate out three characteristic regimes:
(i) the low temperature region, the physics of which is defined by phononic behaviour of the quasiparticles, leading to a superfluid density which depends on temperature as $(1-\alpha T^3)$; (ii) a free particle region, where $\rho_s$ behaves linearly with temperature; (iii) and a critical region, determined by the fluctuations around the critical temperature $T_c$ ( $\rho_s$ vanishes at $T_c$) (see Fig.6.) 

\begin{figure}[!ht]
\begin{center}
\epsfxsize=0.45\textwidth{\epsfbox{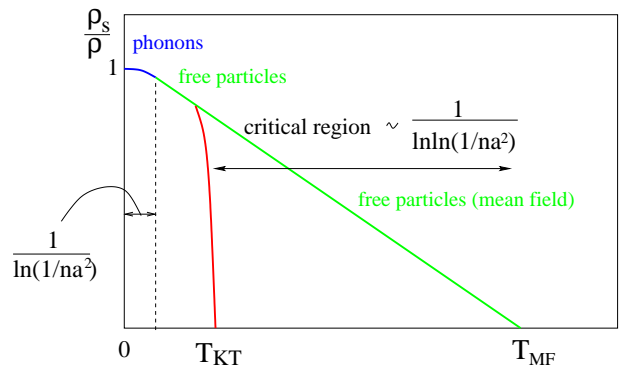}}
\end{center}
\caption{Schematic phase diagram of a uniform dilute Bose gas. $\rho_s$ is the superfluid density, normalized by the mass density of the gas $\rho=mn$, $T_{KT}$ is the critical temperature of the Kosterlitz-Thouless transition, $T_{MF}$ is the mean-field temperature, calculated perturbatively. Size of the transition critical region is defined by a parameter $1/\ln\ln (1/na^2)$, where $a$ is characteristic s-wave scattering length. The region, dominated by phononic quasi-particle behaviour is of the width $1/\ln(1/na^2)$.}
\label{sf}
\end{figure}

The diluteness criterion is determined by the condition that the critical region is small enough so that all three regimes can be well separated. The width of the first regime is in fact given by Schick's small parameter (\ref{dilute_s}), while the size of the critical region is characterized by the double log (\ref{dilute_FH}). The problem however, is that for all practically relevant situations, even for very small $na^2$, Fisher and Hohenberg's small parameter  $1/lnln(1/na^2)$ is still orders of magnitude greater than $1/ln(1/na^2)$. This means that in practice the critical region associated with Kosterlitz-Thouless transition is so large, that mean-field based approaches do not give any reliable results. Note, that the double log result was also reproduced by Fisher and Hohenberg  in a more accurate way within a renormalization group treatment of the same problem.  They have also estimated the superfluid transition temperature, which reads 
\begin{equation}
T_c\sim \frac{2\pi n}{m \ln \ln [1/(na_s)^2]}.
\label{crittemp}
\end{equation}

The results of FH work \cite{Fisher:1988} have been confirmed in other approaches, see for example virial expansion of a dilute Bose gas by \textcite{Ren:2004} or RG analysis by \textcite{Kolomeisky:1992a,Kolomeisky:1992b} and by \textcite{Crisan:2001}.  \textcite{Pieri:2001a,Pieri:2001b} demonstrated by analyzing the normal state by standard diagram technique that
 the transition temperature (\ref{crittemp}) appears as a lower bound for the validity of the $t$-matrix as a controlled approximation for the dilute Bose gas.

The FH diluteness condition (\ref{dilute_FH}) is extremely stringent, and if straightforwardly applied to experimentally relevant situations  \cite{Goerlitz:2001,Rychtarik:2004}, would  mean that the systems observed to undergo  a BEC phase transition in 2D are not actually dilute, and could never be so.
This line of reasoning motivated Liu and Wen \cite{Liu:2002} to come up with an exotic alternative scenario involving a two-dimensional strongly-correlated spin liquid. 

The extreme conclusions drawn from the FH diluteness criterion are nevertheless related to the general drawbacks of the Popov approximation. We will see in the next section, that in a more realistic model, which takes into account interactions in a self-consistent way, the diluteness condition becomes much weaker.  Moreover, in view of our previous discussion about the inapplicability of arguments based on  homogeneous systems in the thermodynamic limit to trapped gases, it would seem that the FH diluteness requirement is not really relevant for the experimental situation of the Bose gas in a magnetic trap.

\subsection{Other approaches: RPA, Many-body $t$-matrix, Monte Carlo} \label{sec_MBT}

In this section we review a range of diagrammatic approaches that have built upon the early RPA and $t$-matrix approximation in order to improve the description of  the superfluid of BKT transition and also the numerical methods, which allow to directly probe  the critical region of the 2D transition.

The first finite temperature generalization of Bogoliubov random phase approximation (\ref{RPA_Bog}) was introduced by \textcite{Tserkovnikov:1964} (english version \textcite{Tserkovnikov:1965}). His concern was to calculate the finite-temperature correction to the condensate density in 3D dilute Bose gases with weak interactions. Tserkovnikov assumed that the average single particle kinetic energy is small, compared to the potential energy for all temperatures below $T_c^{3D}$. He also remarks that his approximation does not meet the Landau superfluidity criterion and that more precise equations should be sought in future work.  

The RPA method was further developed in the papers of \textcite{Szepfalusy:1974}, whose main interest was the investigation of dynamics of the second order phase transition. Around the same time a large-$N$ approach was applied to the Bose gas by \textcite{Abe:1974d} and \textcite{Abe:1974}, who calculated the dynamical scaling for one-particle Green's function up to $O(1/N)$. 
Here the idea of the large-$N$ approach is to expand the number of independent components of the Bose field from unity to $N/2$ using $1/N$ as an expansion parameter. To produce a controlled large $N$ limit, the interaction strength is scaled to be of order $1/N$.

The RPA large-$N$ method has been applied to 2D Bose gas by \textcite{Nogueira:2006}. The interaction in their approach is approximated by a 2D coupling constant, derived in the $t$-matrix approximation, considered by Popov and Schick (see sections \ref{sec_Beliaev}, \ref{sec_Popov} and \ref{sec_dilute}). It is however known that in a large-$N$ approach one can not simultaneously account for both particle-hole channel (RPA) and particle-particle channel in a well-controlled fashion. 
Nevertheless, the authors claim that the diluteness condition leads to the appearing of $t$-matrix diagrams first, while the next class of diagrams are those from particle-hole channel \cite{Nogueira:2006}. This approximation results in the Bogoliubov quasiparticle dispersion containing a log correction due to low dimensionality  $\xi_k=\sqrt{\epsilon_k^2+2g_{2D}n\epsilon_k\left[1-\frac{Tm}{\pi n}ln(ka)\right]}$, so that excitations in the system exhibit a roton-like minimum. Note that the excitation spectrum is calculated under the assumption that one-particle Green's function and density correlators share the same poles (this property was derived by \textcite{Hohenberg:1965} in case of 3D condensed Bose system). Would be interesting to check if these RPA results are confirmed in other approaches. 

We now proceed to a discussion of the various generalizations of the two-body $t$-matrix approach. 
Though simple and elegant, the perturbative two-body $t$-matrix approach does have its drawbacks. 
The main problem is related to its inability to properly describe the critical region in low dimensions. 
For example, the $t$-matrix method does not predict the Nelson-Kosterlitz universal jump in the superfluid density. In 3D the two-body $t$-matrix approach leads to a first order phase transition for the condensate density, which is the consequence of non-self-consistency of this first order perturbative approximation (see also \textcite{Griffin:1988,Lee:1958,Reatto:1969}). 

Many of these problems can be solved if  the many-body corrections, arising due to the surrounding gaseous medium are taken into account. This is the key idea in the  ``many-body $t$-matrix approximation'' (see the comprehensive review by \textcite{Shi:1998}, low-dimensional systems within many-body $t$-matrix approach are analyzed in the papers by \textcite{Stoof:1993,Andersen:2002bc,Al Khawaja:2002}, for Hartree-Fock -Bogoliubov study of a two-dimensional gas see recent works by \textcite{Gies:2004a,Gies:2004b,Gies:2005}).

Since the  ``many-body $t$-matrix'' methods are extensively discussed in the literature, here we restrict ourself to a brief description of the method, providing all relevant references. The Bogoliubov-Hartree-Fock (BHF) approximation (see \textcite{Griffin:1996} and the analysis of excitations in a trapped 3D gas paper by \textcite{Hutchinson:1997}) has a Heisenberg equation of motion for a Bose field operator of the kind (\ref{wfGP}) as a starting point
\begin{eqnarray}
i\hbar\frac{\partial}{\partial t}\psi(r,t)&=&\left(-\frac{\hbar^2}{2m}\bigtriangledown^2+V_{ext}({\bf r})-\mu\right)\psi(r,t)+ \nonumber \\
&+&g\psi^+(r,t)\psi(r,t) \psi(r,t)
\label{BHF1}.
\end{eqnarray}
 A short-range interaction is assumed among the atoms $U({\bf r-r'})=g\delta({\bf r-r'})$. 
Treating the interaction term in this equation in the self-consistent mean-field approximation, one arrives at the  equation
\begin{eqnarray}
&&\left(-\frac{\hbar^2}{2m}\bigtriangledown^2+V_{ext}({\bf r})-\mu\right)\phi(r)+
\nonumber \\
&&+
g ( n_0(r)+2n'(r))\phi(r) +
g m'(r) \phi^*(r) =0 \qquad
\label{BHF2},
\end{eqnarray}
where $n_0$ is the condensate density, $n'(r)=\langle \psi'^+(r)\psi'(r) \rangle$ and $m'(r)=\langle \psi'(r)\psi'(r)\rangle$ (anomalous average). In order to describe excitations in the system one should also write down the equation of motion for $\psi'$
\begin{eqnarray}
i\hbar\frac{\partial}{\partial t}\psi'(r)&=&\left(-\frac{\hbar^2}{2m}\bigtriangledown^2+V_{ext}({\bf r})-\mu\right)\psi'(r,t)+ \nonumber \\
&&+ 2g n(r)\psi'(r,t)+gm(r)\psi'^+(r,t)
\label{BHF3}
\qquad
\end{eqnarray}
(with $n(r)=n_0+n'$ and $m(r)=\phi^2+m'$). Equations (\ref{BHF2}) and (\ref{BHF3}) correspond to the Bogoliubov-Hartree-Fock approximation: a bosonic analog of the finite-temperature Bogoliubov-de Gennes equations. These equations can also be re-expressed in terms of a Green's function formalism. Note, that the appearance of anomalous averages in BHF formalism  (which are not present in Popov approach), leads to a gap in the quasiparticle excitation spectrum.   

Many body effects are also effectively treated in a variational method, applied to dilute Bose gases by  \textcite{Bijlsma:1997}, and many-body $t$-matrix methods which have been significantly improved in recent years (see \textcite{Andersen:2002bc,Al Khawaja:2002}). A time dependent BHF approximation has recently been developed by \textcite{Proukakis:1998}. Here, the authors \cite{Proukakis:1998} claim that the pseudopotential approximation $U({\bf r-r'})=g\delta({\bf r-r'})$ should be imposed only after the effective interaction is expressed in terms of many-body $t$-matrix. Both approaches, the one used by Griffin et al. and another, developed by Stoof and collaborators, are qualitatively similar, in that they treat interactions in many-body $t$-matrix approach, but they differ in some details, for instance  in selecting  out the important diagrams. 

Let us now  discuss some of the results of the many-body $t$-matrix method for 2D systems. More than a decade ago \textcite{Stoof:1993} demonstrated that the infrared divergences, appearing in the two-body $t$-matrix treatment, can be elegantly eliminated when the effects of surrounding gas are taken into account. With this approach the universal jump in superfluid density, predicted by Nelson and Kosterlitz can be reproduced. 
A weaker diluteness condition, namely that of Schick (\ref{dilute_s}), defines the applicability of this many-body $t$-matrix approximation, one that is satisfied experimentally in systems, such as  spin polarized  atomic hydrogen, absorbed on $^4$He surface \cite{Stoof:1993}.

The conclusions of \textcite{Petrov:2000,Petrov:2001} have been confirmed in recent investigations by \textcite{Andersen:2002bc} and \textcite{Gies:2004a,Gies:2004b,Gies:2005}.   
 The theory of \textcite{Andersen:2002bc}, free  of infrared divergences in all dimensions, allows for calculation of the density profile of a (quasi)-condensate cloud of a gas for any aspect ratio of the trap (within local density and Thomas-Fermi approximation). At very low temperature, depending on the trapping geometry, the presence of  a true condensate in the equilibrium state is found. Hutchinson and coworkers \cite{Gies:2004a,Gies:2004b,Gies:2005} also ``see'' within their HFB approach  
 a macroscopic occupation of the ground state at low temperatures, implying the presence of a condensate state.

 To conclude, the presence of the trap appears to stabilize the condensate against the long-wavelength fluctuations and the BEC state can form at finite, though very low temperatures, when the discrete nature of the energy spectrum is taken into account.

Most reliable description of the 2D Bose gas to date is provided by Monte Carlo simulations \cite{Kagan:2000,Prokof'ev:2001,Prokof'ev:2002}, because it allows to study the critical region of the BKT transition, which is effectively very large and therefore unaccessible to perturbative methods. 
The numerical analysis is simplified, by the fact that the critical properties of all XY models are the same (see \ref{LRO2D}). It suffices therefore to study  the classical $|\Psi|^4$ model on the lattice within a Monte Carlo algorithm. 

Consider, for instance, the temperature-dependence of the particle density in the critical region of the BKT transition, which follows from perturbative analysis \cite{Popov:1983,Kagan:1987,Fisher:1988} of weakly-interacting system
\begin{equation}
n=\frac{mT}{2\pi}\ln\frac{C}{mU_{eff}},
\end{equation} 
where $U_{eff}$ is an effective interaction, proportional to $f_0^{2D}$ \eqref{t_schick}, and $C$ is the constant, which is not possible to evaluate within perturbative expansion in powers of $U_{eff}$ \cite{Prokof'ev:2001}. Monte Carlo estimation  gives  $C=380\pm 3$; this large value of $C$ makes it virtually impossible to reach the limit of small $U_{eff}$ for weakly-interacting system.

At transition we obtain an an accurate microscopic expression for the critical temperature of BKT transition \cite{Prokof'ev:2001,Prokof'ev:2002}.
\begin{equation}
\frac{T_{KT}}{n}=\frac{2\pi}{m\ln(C/mU_{eff})}.
\end{equation}  
It is interesting to compare this density $n$ to the quasi-condensate density $n_q$ and superfluid density $n_s$ in the critical region. It turns out, that $n_q/n$ is of order of unity, unless $mU_{eff}$ is exponentially small, while the ratio $n_q/n_s$ is of order of 2, which means that superfluid density is substantially smaller than quasi-condensate density at the transition.

The temperature behaviour of various densities, obtained in Monte Carlo procedure, can be used for checking whether RG and perturbative approaches essentially overlap. Indeed, 
Monte Carlo simulations have been able to capture the crossover between the mean-field behaviour and the critical fluctuation region described by the KT transition \cite{Prokof'ev:2002}. \textcite{Prokof'ev:2002} show that this crossover is characterized by a universal ratio of the superfluid and quasi-condensate density. One can also see that the conventional mean-filed result $n_s/n\sim 1-T/T_{KT}$ is not valid anywhere, while the modified mean-field theory introduced by \textcite{Prokof'ev:2002} can predict accurately the behaviour of the quasi-condensate density up to $T_{KT}$. 

\subsection{Breathing modes of 2D systems} \label{sec_modes}

At the end of this chapter we consider a universal property of a two-dimensional gas with a {\it contact} interaction, confined in a harmonic potential. 
  \textcite{Pitaevskii:1997} predict that such a system develops oscillations or breathing modes, which can be probed  experimentally or in simulations and thus can serve as a practical criterion of the two-dimensional nature of a system. 

   The appearance of  breathing modes is related to a hidden ``Lorentz'' symmetry inherent to any two-dimensional Hamiltonian of the following general form
\begin{equation}
H=H_0+H_{ext}
\label{Ham}
\end{equation}
where
\begin{equation}
H_0=\sum_i \left(-\frac{1}{2m}\triangle_i\right)+\sum_{i<j}U({\bf r}_i-{\bf r}_j),
\end{equation}
and $H_{ext}=\sum_i\frac{1}{2}m\omega_0^2r_i^2$ is a harmonic potential.
  
It is readily seen that $H_0$ is scale invariant in case of a local 2D  interaction 
\begin{equation}
U({\bf r}_i-{\bf r}_j)=\frac{g}{2}\delta^2({\bf r}_i-{\bf r}_j),
\label{potential}
\end{equation}
(in fact it is scale invariant for any potential with a property $U(l{\bf r})=U({\bf r})/l^2$).
The presence of a trap breaks the scale invariance of $H_0$. Note, that in principle the scaling invariance of the Hamiltonian $H_0$ is broken in 2D, because then scattering phase shift is energy dependent due to the logarithmic dependences characteristic of two dimensions (phase shift is proportional to the coupling constant $g^{2D}$ or to $1/ln(ka)$). The energy-dependent phase shift signals the breaking of scale invariance at the quantum level \cite{Cabo:1998}. But this symmetry breaking is explicit and is not attributed to any phase transition physics. 

 In spite of the breaking of scale invariance,  because of a special property of the harmonic oscillator, a powerful spectrum generating symmetry still exists. That can be seen from the commutator $[H_{ext},H]=i\omega_0^2Q$, where $Q=1/2\sum_i({\bf p}_i{\bf r}_i+{\bf r}_i{\bf p}_i)$ is the generator of scale transformations. One can check that $[Q,H_0]=2iH_0$ and $[Q,H_{ext}]=-2iH_{ext}$. These results can be formulated within the well-known algebra of SU(1,1) or SO(2,1) symmetry group, i.e. the 2D Lorentz group.

Starting from the lowest energy state $\epsilon_0$ one can produce higher order states with energies $\epsilon_0+2n\omega_0$ ($n=1,2,...$) by applying one of the SO(2,1) group generators  $L^+= (L_1+i L_2)/\sqrt 2$ where $L_1=(H_0-H_{pot})/2\omega_0, L_2=Q/2$ (the correspondent annihilation operator is $L^-=(L_1-i L_2)/\sqrt 2 $). The excitations with the energies $2n\omega_0$ are associated with the breathing, or pulsating modes of the system. 

As an example, the authors \cite{Pitaevskii:1997} considered the classical Gross-Pitaevskii equation and predicted the existence of undamped breathing modes in the condensate.
The appearance of transverse breathing modes with a frequency equal to an integer multiple of the trap oscillation frequency was observed experimentally in  an elongated condensate of $^{87}Rb$ atoms \cite{Chevy:2002}.
Numerical simulations (exact diagonalization) seem to indicate the existence of dipole or breathing modes in a 2D system even for relatively small number of atoms \cite{Haugset:1998}.

BHF study of a 2D Bose gas by Gies {\it et al.}  \cite{Gies:2004a,Gies:2004b,Gies:2005} has also showed that at low temperatures the frequency of the lowest lying excitation ($n=0$ mode) is precisely $2\omega_0$, independent of the interaction strength.  At high temperatures the frequency of this mode shifts to a lower frequency region, being modified by the addition of the potential from the static thermal cloud.

Nevertheless, it is important to note that a $\delta$-function is not well-defined in two dimensions due to logarithmic UV divergences \cite{Pitaevskii:1997}, that are cut off by the finite range of interaction, whether it is a small or large effect, should be investigated.

\section{Conclusions and open questions} \label{Conclusions}

We have surveyed a number of theoretical issues arising in the field of a weakly-interacting uniform or confined in a trap dilute Bose system at low temperatures in 2D. 
The underlying physics of such a system depends on the size of the system, degree of its inhomogeneity, and temperature. 

If the system is very large and uniform, one might expect realization of the BKT transition,  characterized by the presence of a topological order below the critical temperature $T_{KT}$ down to zero temperature when the true long-range order (BEC) forms. Perturbative approaches, based on a low density approximation and point-like or short-range interactions, surveyed in this Colloquium, are not really suited to a description of the vortex excitations of the ordered phase of BKT transition, however, these methods provide a good description of many physical properties. For example, a modified version of the mean-field theory, the ``many-body $t$-matrix approach'' is able to capture the second order nature of the phase transition in 3D and the Nelson-Kosterlitz universal jump of superfluidity in 2D.

In experiment the low temperature regime of a quantum gas is achieved by confining the atomic system in an external potential. The proper description of a practically realized system requires therefore the inclusion of the inhomogeneities, introduced by a trap.  In 3D the effect of the trap is not very pronounced, and it is possible to calculate the correction to the critical temperature due to interactions perturbatively \cite{Arnold:2001}, while in the uniform case the RG study is required \cite{Ledowski:2004} (for details see the book by \textcite{Pitaevskii:2003}).

In 2D, as we have seen in the case of a gas without interactions, the presence of the trap dramatically modifies properties of the system (density of states), so that BEC becomes possible at finite temperature. The inclusion of interactions into the picture is a complicated task. First of all, the system is inhomogeneous and all previously developed perturbative methods, such as the $t$-matrix approach are strictly speaking not suitable for its analysis. In principle, one should solve the many-body scattering problem in a trap, taking into account the discrete spectrum and the finite range of a potential, which is extremely difficult.  One can of course consider a simplified problem of a quasi-homogeneous trap and adapt well-studied techniques for  that case.  As we have seen, the main effects of the trap are indeed captured at least qualitatively within such a scheme.

Intuitively, it is clear, that inhomogeneities would tend to suppress the universal jump of superfluidity, and would rather favor the ``true'' BEC state at low temperatures in a system with finite number of particles (because the long-range thermal fluctuations will be quenched by a trap). 

Very recently \textcite{Holzmann:2005} have analyzed the behaviour of the weakly-interacting trapped system {\it in the thermodynamic limit} within local density approximation (LDA). They have shown, that although the universal Nelson-Kosterlitz jump is indeed not present, the system does undergo BKT transition at the temperature, somewhat lower than $T_c^{2D}$ \eqref{tc2d}.

 In a  paper by \textcite{Simula:2005} the authors have predicted that both a BE condensed phase and a KT superfluid phase, separated by a first order transition, will be present in a 2D trapped gas. 
The authors arrive at their conclusions by comparing the Helmholtz free energy of the ground state, characterized by a condensate wave-function and an excited state, containing a vortex pair in the ordering field. Based on entropy considerations, they find a critical temperature $T_{c}$ above which a thermally activated transition to the state containing vortex-pair excitations becomes favourable. These conclusions require, however, solid confirmations from both theory and experiment.

A challenging  problem is however the actual estimation of the BEC transition temperature in a 2D interacting system and the relationship of the quasi-condensate state to the KT vortex-pair plasma state, depending on the geometry of the system and  number of particles. It is also important to develop numerical methods adapted for experiment.  One of the promising developments in this direction has been made by \textcite{Davis:2001a,Davis:2001b,Davis:2002}, who have proposed a projected Gross-Pitaevskii equation formalism, which allows the efficient investigation of  finite temperature properties of the equilibrium condensate state - even in the region where the Bogoliubov theory fails. An extension of this method to harmonically trapped condensates has been considered in the recent paper by \textcite{Blakie:2005}. Just this year \textcite{Simula:2006} have analyzed the 2D Bose gas by classical field methods, adopted from quantum optics. They have demonstrated that two distinct superfluid phases, separated by thermal vortex-antivortex pair creation, exist in experimentally producible quasi-2D Bose gas. \textcite{Simula:2006} provided a strong evidence that a strange (``zipper'') interference pattern observed in a recent experiment by \textcite{Stock:2005} can be explained by a presence of a vortex excitations in an experimental system.   

One can thus tentatively identify the following phases in a trapped 2D Bose gas (Fig.7):
\begin{figure}[!ht]
\begin{center}
\epsfxsize=0.45\textwidth{\epsfbox{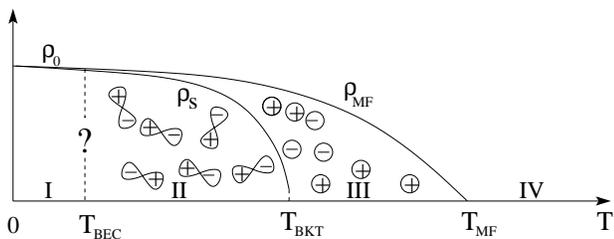}}
\end{center}
\caption{Schematic phase diagram of a 2D trapped weakly-interacting dilute Bose gas. $\rho_0$ stands for the density of  the condensate, $\rho_s$ is the superfluid density, $\rho_{MF}$ is the mean-field density, which can be estimated perturbatively. $T_{BEC}$ is the crossover temperature from the superfluid regime to the true Bose-Einstein condensation phase. $T_{BKT}$ is the critical temperature of the Kosterlitz-Thouless transition. $T_{MF}$  marks the critical region of BKT transition.}
\label{tmatrix}
\end{figure}

\begin{itemize}

\item phase I: low -temperature true BEC phase;

\item phase II: KT vortex-antivortex pair superfluid, or quasi-condensate, or condensate with a fluctuating phase, at the transition $T_{KT}$ superfluid universal jump is almost suppressed;

\item phase III: critical region of the BKT transitiona; vortex pair dissolve, above $T_{KT}$ vortices are unbound and free;

\item phase IV: above the mean-field temperature it is a normal fluid, no local order parameter or vortices exist.

\end{itemize}

The phase diagram of a real system will depend on many factors, as we stressed at the beginning of this Chapter. Many aspects of the diagram, depicted on Fig. 7 require careful investigation, and reliable confirmation from both theory and experiment. 

In ending we attempt to summarize some of the open questions:
\begin{itemize}

\item Nature of superfluid phases of a 2D weakly-interacting Bose gas: what  is the nature of the crossover to a superfluid phase?  What is the explicit relation between  superfluidity and the ``quasi-condensate'' state? Is there a crossover to BEC state at low temperatures? and if yes, under which conditions?

\item Can we help experimentalists to ``see'' the vortex excitations in the superfluid state, to really identify the KT state? There is a clear need in good vortex detection methods. Can, for example, disorder help us to pin the vortices?  

\item What are in general measurable physical properties, which  delineate between the coherent condensed state and superfluidity? Initial progress in this direction has been already made, \textcite{Polkovnikov:2005} have suggested how to identify the KT transition from experimentally measured interference pattern.

\item Can we solve the many-body scattering problem in the trap? Does diluteness of the gas simplify this problem?

\item Is local density a good approximation for description of the experimental systems, and if yes, under which conditions?

\item Could one justify a large $N$ approach which improves on the existing method by incorporating the $t$-matrix approximation?

\end{itemize}

We have also discussed the diluteness condition, derived by \textcite{Fisher:1988} under certain conditions of the transition to superfluid state. In a finite-sized experimental system this condition is not really applicable, and one should use the criterion of Schick \eqref{dilute_s}, which can be seen from the analysis of quantum fluctuations of the 2D BEC at zero temperature \cite{Petrov:2004}.

We have considered only low density approximations.  When the gas is dense, approaches such as that of Gross and Pitaevskii are not applicable. In such high density regimes, new ``strong coupling'' approaches are required. One of the possible solutions may be the slave-boson approximation, which is valid for hard-core bosons at any density \cite{Ziegler:1997}. 

We have not discussed in our Colloquium the role of disorder in the continuum Bose system, though it could be a subject of a separate review and opens up a lot of interesting perspectives. Recent Monte Carlo studies  predict, for example, that  for strong disorder the system enters an unusual regime, where the superfluid fraction is smaller than the condensate fraction \cite{Astrakharchik:2002}.
  Weak disorder can be treated within Bogoliubov theory \cite{Huang:1992,Giorgini:1994} and the striking result of this is that disorder is more active in reducing superfluidity than in depleting the condensate. These results suggest that the relation of superfluidity and Bose-Einstein condensation require further theoretical and experimental investigation.

One can not but mention the rapidly increasing interest to the cold gases with dipole-dipole interactions, which are responsible for a variety of novel phenomena in ultracold dipolar systems, see, for instance,  \textcite{Santos:2003,Santos:2000,Pedri:2005,Stuhler:2005,Fischer:2006} and references therein.

Finally, the problem of measurable quantities is in fact one of the most important in the context of trapped Bose gases.
Unlike an electron system, one cannot attach ``leads'' to the system and measure transport properties of a condensate cloud. One of the most pressing practical needs for theorists and experimentalists is therefore the development of controllable new methods to probe the trapped condensate.

\section*{Appendix I (for section \ref{LRO2D})}

{\bf Sketch of the derivation of Mermin-Wagner-Hohenberg theorem}

One should use the Bogoliubov  inequality
\begin{equation}
\frac{1}{2}\langle \{A,A^+\}\rangle \langle[[C,H],C^+] \rangle \geq T|\langle [C,A] \rangle|^2,
\label{Bogoliubov_in}
\end{equation}
where the average $\langle X \rangle=Tr \left(X e^{-\beta H}\right)/Tr e^{-\beta H}$, and operators $A$ and $C$ are such, that the ensemble averages on the lhs of (\ref{Bogoliubov_in}) exist. The inequality (\ref{Bogoliubov_in}) follows quite straightforwardly from the Schwartz inequality
\begin{equation}
(A,A)(B,B)\geq |(A,B)|^2
\end{equation}
where a scalar product is defined by $(A,B)=T\int \frac{d\omega}{\pi}\frac{1}{\omega}\chi_{AB}(k\omega)$, where $\chi_{AB}$ is the Fourier transform of the response function $\chi_{AB}(rt,r't')=\langle \frac{1}{2\hbar}[A(rt),B(r't')]\rangle$ (see for example the textbook by \textcite{Forster:1990}).

In the case of a Bose system the derivation of \textcite{Hohenberg:1967} is based on Bogoliubov and Schwartz inequalities and the $f$ {\it sum rule}
\begin{equation}
T\int \frac{d\omega}{\pi}\frac{1}{\omega}\chi_{AA^+}(-k\omega)=\frac{k^2n}{m}.
\label{sumrule}
\end{equation}

{\bf Derivation by Fischer of geometrical analog of the Hohenberg inequality}

The bosonic field operator is as usual decomposed into condensate and noncondensate parts
\begin{equation}
\psi({\bf r})=\psi_0({\bf r})a_0+\delta \psi({\bf r}).
\label{op_uwe}
\end{equation}
The key observation of \textcite{Fischer:2002} is that the Bogoliubov prescription should be  applied {\it after} implementing the commutation relation
\begin{equation}
[\delta \psi({\bf r}),\delta \psi^+({\bf r'})]=\delta({\bf r}-{\bf r'})-\psi_0({\bf r})\psi_0^*({\bf r'}),
\label{comm}
\end{equation}
for otherwise, the second term on the rhs of (\ref{comm}), which turns out to be crucial for calculating the condensate fraction correctly, would vanish. 

The operators $A$ and $C$ in the Bogoliubov inequality (\ref{Bogoliubov_in}) are chosen to be smeared excitation and total density operators
\begin{equation}
\hat A=\int d^dr f_A({\bf r}) \delta \psi({\bf r}); \quad
\hat C=\int d^dr f_C({\bf r}) \delta \rho({\bf r}),
\end{equation}
where $f_A$ and $f_C$ are carefully chosen ``smearing functions'' ($f_C({\bf r})\sim \psi_0^*({\bf r})$, $f_A({\bf r})\sim e^{i\bf k r}$). Next, the $f$-sum rule analogous to (\ref{sumrule}), can be derived in coordinate space. 

\section*{Appendix II (for the section \ref{sec_Popov})}

{\bf Popov's approach}

 To derive the phase transition curve for a two-dimensional interacting Bose gas, one needs to explore the finite-temperature behaviour of the $t$-matrix (\ref{tmatrix2}). The Bethe-Salpeter equation (Fig.3) in the Matsubara representation reads
\begin{widetext}
\begin{eqnarray}
\Gamma(p_1,p_2;p_3,p_4)=U_{{\bf k_1}-{\bf k_3}}-\frac{1}{\beta V}\sum_{q,i\omega_l}U_{\bf q}G^0({\bf k_1}-{\bf q},i\omega_1-i\omega_l) 
\nonumber \\ \times  
G^0({\bf k_2}+{\bf q},i\omega_2+i\omega_l)\Gamma(p_1-q,p_2+q;p_3,p_4),
\label{gamma_T}
\end{eqnarray}
\end{widetext}
where  $\omega_j=2\pi jT$ is an even Matsubara frequency, and  the four-dimensional vector $p_j\equiv ({\bf k}_j,\omega_j)$ represents the momentum ${\bf k}_j$ and frequency $\omega_j$ of the particle before scattering $(j=1,2)$ and after $(j=3,4)$. Energy and momentum conservation requires $p_1+p_2=p_3+p_4$.

The main contribution to the sum over internal momenta in (\ref{gamma_T}) comes from $k \sim a^{-1}$ which is due to the potential, discussed above. Since $a^{-1} \gg \sqrt T \gg \sqrt \mu$,  the $\mu$ dependence in the Green's function can be safely neglected \cite{Popov:1983}. Consequently, after integrating over frequencies Eq. (\ref{gamma_T}) is reduced to a $t$-matrix equation  
\begin{eqnarray}
t(p_1,p_3,z)&+&\int \frac{dp'}{(2\pi)^2}U(p_1-p')\frac{1}{p'^2/m-z}t(p',p_3,z)=
\nonumber \\
&=&U(p_1-p_3)
\label{tmatrix_T}
\end{eqnarray}

Schematically, the $t$-matrix equation can be expressed as $t_z+UR_zt_z=U$, with $R_z=1/(p^2/m-~z)$. The operator $(1+UR_z)$ can be inverted and we get $U=t_z(1-R_zt_z)^{-1}$. In this fashion the interaction is eliminated from the $t$-matrix equation and we arrive at the Hilbert identity $t_z-t_{z_0}=t_z(R_{z0}-R_z)t_{z_0}$. This last equation is readily integrable, since at low energies ($p_1,p_2\ll 1/a, |z|\ll 1/(ma^2)$) we can neglect the momentum dependence of the $t$ matrix. The energy $z_0$ defines an  arbitrary high-energy cut off of the order $1/(ma^2)$, so that $t(z_0)\gg t(z)$, and we obtain the long-wavelength asymptotic of the $t$-matrix in 2D
\begin{equation}
t\simeq \frac{4\pi}{m \ln [\epsilon_0/(-z)]}
\label{lowetmatrix}
\end{equation}
We see that in 2D, the $t$-matrix vanishes in the limit $p_1,p_2,z\rightarrow 0$ and in fact diverges at the high-energy cut off $\epsilon_0=|z_0|$.

Next we need to integrate out the high-energy modes with momenta $k>k_0'$ in our functional $\exp S$  (\ref{action}). The cut-off $k_0'$ is defined as
\begin{equation}
max(|\mu|,T)<< \frac{(k_0')^2}{2m}<<\epsilon_0
\end{equation}
The result of this integration is the reduced action
\begin{eqnarray}
&&S'=\sum_{\omega, k<k_0'}\left( i\omega-\frac{k^2}{2m}+\mu \right)a^+(p)a(p)- 
\nonumber \\
&&-\frac{1}{2\beta V} \sum_{p_1+p_2=p_3+p_4} t'a^+(p_1)a^+(p_2)a(p_3)a(p_4)
\qquad
\end{eqnarray}
where all the summations are cut-off at $k=k_0'$ and {\it the potential is replaced by a $t$-matrix} with
\begin{equation}
t'=t'(\omega)=\frac{4\pi}{m \ln (\epsilon_0/[(k_0'^2/m)-i\omega])}
\end{equation}
Now the functional $\exp S'$  is to be integrated over the variables $a^+(p),a(p)$ within the momentum shell $k_0<k<k_0'$, where $k_0$ is defined from the inequality $k_0^2/m>>T/\ln |\epsilon_0/\mu|$ and serves to distinguish between slow and rapid particles. Variables with momenta smaller than $k_0$ are taken into account in the action by the transformation
\begin{equation}
a^+(p),a(p)\rightarrow (\rho_0(k_0)\beta V)^{1/2}\delta_{p0}
\label{transform}
\end{equation}
here $\rho_0(k_0)$ is the density of slow particles and one can introduce the density of fast particles $\rho_1(k_0)$.

After the transformation (\ref{transform}) one can make use of  standard perturbation theory formalism and derive expressions for the densities $\rho_0$ and $\rho_1$. Their sum gives the total density  $\rho=\rho_0+\rho_1=\rho_n+\rho_s$, which is independent of auxiliary momenta $k_0$ and $k_0'$
\begin{equation}
\rho=\frac{m\mu}{4\pi}(\ln \epsilon_0/\mu -1)-\frac{1}{(2\pi)^2}\int d^2k \frac{k^2}{2m\epsilon(k)}n_B(k)
\label{totaldensity}
\end{equation}
where $n_B(k)=(e^{\beta\epsilon_k}-1)^{-1}$, and the formulae for the normal and the superfluid component densities read
\begin{eqnarray}
&&\rho_n=\frac{\beta}{(2\pi)^2}\int d^2k \frac{k^2}{2m}n_B(k)(1+n_B(k)) \cr
&&\rho_s=\frac{m\mu}{4\pi}(\ln \epsilon_0/\mu -1)- \nonumber \\
&&-\frac{1}{(2\pi)^2}\int \frac{d^2k}{2m} k^2 n_B(k)\left[\frac{1}{\epsilon_k}-\beta (n_B(k)+1)\right].
\qquad
\label{densities}
\end{eqnarray}
Eqs (\ref{totaldensity}) and (\ref{densities}) define the thermodynamics of the system below the phase transition. Above the phase transition one can use the standard perturbation theory for calculation of the Green's functions and the thermodynamical quantities. The critical temperature is derived now from the condition $\rho=\rho_n$ at the transition.

\acknowledgements
I would like to acknowledge discussions with P.B. Blakie, A. Chubukov,  D. Efremov, M. Garst, M. Greiter,  A. Rosch and P. Woelfle. I highly appreciate many insightful discussions with M. Eschrig, U. Fischer and F. Nogueira. I am indebted to P. Coleman for reading the manuscript and numerous critical comments. I acknowledge the Humboldt foundation for support and Kavli Institute for Theoretical Physics of Santa Barbara for hospitality, under NSF grant PHW 99/07949.

\bibliographystyle{apsrmp}

\newpage

\begin{table}
\caption{Properties of the ideal gas. ``DOS'' stands for density of states $\rho$, ``TDL'' stands for thermodynamic limit. $T_{BEC}$ is the critical temperature of Bose-Einstein condensation depending on the dimension. $N$ is the number of particles, $V$ is the volume of the system, $\omega_{ho}$ is the geometric average of the oscillator frequencies: in 2D $\omega_{ho}=(\omega_x \omega_y)^{1/2}$ and in 3D $\omega_{ho}=(\omega_x \omega_y \omega_z)^{1/3}$.}
\label{tab_ideal}
\begin{tabular}{|c||c|c|}
\hline
ideal gas property & uniform & trapped \\
\hline
\hline
DOS & $\rho_{3D}\sim \sqrt{\epsilon}$  & $\rho_{3D}\sim \epsilon^2$ \\
 & $\rho_{2D}\sim const$ & $\rho_{2D}\sim \epsilon$   \\
\hline
$T_{BEC}$ & $T_{c0}^{3D}\sim (\frac{N}{V})^{2/3}$   & $T_{c0}^{3D}\sim \omega_{h0}N^{1/3}$  \\
 & $T_{c0}^{2D}\rightarrow 0$ & $T_{c0}^{2D}\sim \omega_{h0}N^{1/2}$ \\
\hline
TDL & $\lim \frac{N}{V}=const$ & 3D: $\lim \omega_{h0}N^{1/3}=const$\\
 &$N\rightarrow \infty$, $V\rightarrow \infty$ & 2D: $\lim \omega_{h0}N^{1/2}=const$ \\
 & & $\omega_{h0}\rightarrow 0, N\rightarrow \infty$ \\
\hline
\hline
\end{tabular}

\end{table}

\end{document}